\documentclass[a4paper,10.5pt]{article}

\usepackage{blkarray}
\usepackage[hyphens]{url}
\usepackage{mathtools,amssymb}
\usepackage[usenames,dvipsnames,table]{xcolor}
\usepackage[titletoc]{appendix}
\usepackage{mathtools,amssymb}
\usepackage[makeroom]{cancel}
\usepackage{graphicx}
\usepackage{epstopdf}
\usepackage{mathrsfs}
\usepackage{enumitem}
\usepackage{stackrel}
\usepackage{float}
\usepackage{booktabs}
\usepackage[normalem]{ulem}  
\usepackage{cancel} 
\usepackage{hyperref}
\usepackage[final]{showkeys} 
\usepackage{longtable}
\usepackage{physics}
\usepackage{empheq}
\usepackage{cite} 
\usepackage{comment}
\usepackage{tikz}
\usetikzlibrary{decorations.pathmorphing}
\usetikzlibrary{decorations.pathreplacing}
\usetikzlibrary{decorations.text}
\usepackage{braket}
\usepackage{fullpage}
\usepackage{subcaption}
\usepackage{pgfplots}
\usepackage{amsmath}
\usepackage{ytableau}
\usepackage{dsfont}

\DeclareMathAlphabet{\mathpzc}{OT1}{pzc}{m}{it}
\DeclareMathSymbol{\mlq}{\mathord}{operators}{``}
\DeclareMathSymbol{\mrq}{\mathord}{operators}{`'}

\allowdisplaybreaks
\allowbreak

\definecolor{refkey}{gray}{0.75}
\definecolor{labelkey}{RGB}{155,48,48}
\renewcommand*\showkeyslabelformat[1]{%
	\fbox{\parbox[t]{0.8\marginparwidth}{\raggedright\normalfont\scriptsize\url{#1}}}}

\usepackage{etoolbox}
\makeatletter
\patchcmd{\hyper@makecurrent}{%
	\ifx\Hy@param\Hy@chapterstring
	\let\Hy@param\Hy@chapapp
	\fi
}{%
	\iftoggle{inappendix}{
		\@checkappendixparam{chapter}%
		\@checkappendixparam{section}%
		\@checkappendixparam{subsection}%
		\@checkappendixparam{subsubsection}%
		\@checkappendixparam{paragraph}%
		\@checkappendixparam{subparagraph}%
	}{}%
}{}{ \errmessage{failed to patch}}

\newcommand*{\@checkappendixparam}[1]{%
	\def\@checkappendixparamtmp{#1}%
	\ifx\Hy@param\@checkappendixparamtmp
	\let\Hy@param\Hy@appendixstring
	\fi
}
\makeatletter

\newtoggle{inappendix}
\togglefalse{inappendix}

\apptocmd{\appendix}{\toggletrue{inappendix}}{}{\errmessage{failed to patch}}
\apptocmd{\subappendices}{\toggletrue{inappendix}}{}{\errmessage{failed to patch}}

\newcommand{\lsim}{\mathrel{\hbox{\rlap{\lower .55ex
				\hbox{$\sim$}} \kern-.3em \raise.4ex \hbox{$<$}}}}
\newcommand{\gsim}{\mathrel{\hbox{\rlap{\lower.55ex
				\hbox{$\sim$}} \kern-.3em \raise.4ex \hbox{$>$}}}}

\begin{document}
	
	
\newcommand{\partiald}[2]{\dfrac{\partial #1}{\partial #2}}
\newcommand{\be}{\begin{equation}}
\newcommand{\ee}{\end{equation}}
\newcommand{\f}{\frac}
\newcommand{\s}{\sqrt}
\newcommand{\lm}{\mathcal{L}}
\newcommand{\wm}{\mathcal{W}}
\newcommand{\om}{\omega}
\newcommand{\bea}{\begin{eqnarray}}
\newcommand{\eea}{\end{eqnarray}}
\newcommand{\ba}{\begin{align}}
\newcommand{\ea}{\end{align}}
\newcommand{\ep}{\epsilon}
\newcommand{\h}{\hat}
\def\ad{a^\dagger}
\def\psid{\psi^\dagger}
\def\ads{AdS$_{\text{2}}$~}
\def\gap#1{\vspace{#1 ex}}
\def\be{\begin{equation}}
\def\ee{\end{equation}}
\def\bal{\begin{array}{l}}
\def\ba#1{\begin{array}{#1}}  
\def\ea{\end{array}}
\def\bea{\begin{eqnarray}}
\def\eea{\end{eqnarray}}
\def\beas{\begin{eqnarray*}}
\def\eeas{\end{eqnarray*}}
\def\del{\partial}
\def\eq#1{(\ref{#1})}
\def\fig#1{Fig \ref{#1}} 
\def\re#1{{\bf #1}}
\def\bull{$\bullet$}
\def\nn{\nonumber}
\def\ub{\underbar}
\def\nl{\hfill\break}
\def\ni{\noindent}
\def\bibi{\bibitem}
\def\vev#1{\langle #1 \rangle} 
\def\mattwo#1#2#3#4{\left(\begin{array}{cc}#1&#2\\#3&#4\end{array}\right)} 
\def\tgen#1{T^{#1}}
\def\half{\frac12}
\def\floor#1{{\lfloor #1 \rfloor}}
\def\ceil#1{{\lceil #1 \rceil}}
	
\def\Tr{{\rm Tr}}
		
\def\mysec#1{\gap1\ni{\bf #1}\gap1}
\def\mycap#1{\begin{quote}{\footnotesize #1}\end{quote}}
		
\def\Red#1{{\color{red}#1}}
\def\blue#1{{\color{blue}#1}}
\def\Om{\Omega}
\def\a{\alpha}
\def\b{\beta}
\def\l{\lambda}
\def\g{\gamma}
\def\ep{\epsilon}
\def\Si{\Sigma}
\def\p{\phi}
\def\z{\zeta}

\def\lan{\langle}
\def\ran{\rangle}

\def\bit{\begin{item}}
\def\eit{\end{item}}
\def\benu{\begin{enumerate}}
\def\eenu{\end{enumerate}}
\def\fr#1#2{{\frac{#1}{#2}}}
\def\gsq{{{\tilde g}^2}}
	
\def\tr{{\rm tr}}
\def\intk#1{{\int\kern-#1pt}}


\newcommand{\com}{\textcolor{red}}
\newcommand{\new}[1]{{\color[rgb]{1.0,0.,0}#1}}
\newcommand{\old}[1]{{\color[rgb]{0.7,0,0.7}\sout{#1}}}
		
\renewcommand{\real}{\ensuremath{\mathbb{R}}}
		
\newcommand*{\Cdot}[1][1.25]{%
	\mathpalette{\CdotAux{#1}}\cdot%
		}
		\newdimen\CdotAxis
		\newcommand*{\CdotAux}[3]{%
			{%
				\settoheight\CdotAxis{$#2\vcenter{}$}%
				\sbox0{%
					\raisebox\CdotAxis{%
						\scalebox{#1}{%
							\raisebox{-\CdotAxis}{%
								$\mathsurround=0pt #2#3$%
							}%
						}%
					}%
				}%
				\dp0=0pt %
				\sbox2{$#2\bullet$}%
				\ifdim\ht2<\ht0 %
				\ht0=\ht2 %
				\fi
				\sbox2{$\mathsurround=0pt #2#3$}%
				\hbox to \wd2{\hss\usebox{0}\hss}%
			}%
		}
		
\newcommand\hcancel[2][black]{\setbox0=\hbox{$#2$}%
\rlap{\raisebox{.45\ht0}{\textcolor{#1}{\rule{\wd0}{1pt}}}}#2} 
		
\renewcommand{\arraystretch}{2.5}%
\renewcommand{\floatpagefraction}{.8}%
	
\def\newthing{\marginpar{{\color{red}****}}}
\def\doubt#1{{\color{red}{[*** #1]}}}
\def\vp{\varphi}
\def\vep{\varepsilon}
\reversemarginpar
		
		
\def\tu{\tau}
\def\ze{z}
\def\d{\partial}
\def\L{\varphi}  
		
\DeclareRobustCommand{\rchi}{{\mathpalette\irchi\relax}}
\newcommand{\irchi}[2]{\raisebox{\depth}{$#1\chi$}}


\hypersetup{pageanchor=false}
			
\vspace{.4cm}
\begin{center}
\noindent{\Large \bf{Exact lattice bosonization of finite $N$ matrix quantum mechanics and $c=1$}}\\
\vspace{1cm} 
Gautam Mandal\footnote{mandal@theory.tifr.res.in}, and 
Ajay Mohan\footnote{ajay.mohan@tifr.res.in}
\vspace{.3cm}
\begin{center}
{\it Department of Theoretical Physics}\\
{\it Tata Institute of Fundamental Research, Mumbai 400005, India.}\\
\end{center}
				
\gap2
\end{center}
\begin{abstract}
We describe a new exact lattice bosonization of matrix quantum mechanics (equivalently of non-relativistic fermions) that is valid for arbitrary rank $N$ of the matrix. It is based on the exact operator bosonization of non-relativistic fermions introduced earlier in \cite{Dhar:2005fg}. The trace identities, which characterize finite rank matrices, are automatically incorporated in the bosonic theory. The finite number $N$ of fermions is reflected in the finite number $N$ of bosonic annihilation-creation operators, and equivalently to the finite number $N$ of lattice points. The fermion Hamiltonian is exactly mappable to a bosonic Hamiltonian. At large $N$, the latter becomes local and corresponds to the lattice version of a relativistic boson Hamiltonian, with a lattice spacing of order $1/N$. The finite lattice spacing leads to a finite entanglement entropy (EE) of the bosonic theory, which reproduces the finite EE of the fermionic theory. Such a description is not available in the standard bosonization in terms of fermion density fluctuations on the Fermi surface, which does not have a built-in short distance cut-off (see, however, a recent description of the finiteness of the {\it fermionic EE} in a collective field theory formalism \cite{Das:2022nxo}). The bosonic lattice constructed in our work is equipped with a geometry that is determined by the matrix potential or equivalently by the shape of the Fermi surface. Our bosonization also works in the double scaled $c=1$ matrix model; in particular the bosonic EE again turns out to be finite, with the short distance cut-off turning into $g_s l_s$, and reproduces the matrix result. Once again, this is to be contrasted with the conventional dual 2D string theory, where the bosonic EE is naturally identified with that of the ``tachyon'', the massless string mode, where one may imagine the short-distance cut-off to be the string length $l_s$. This appears to indicate our bosonization as a different dual description to the $c=1$ matrix model appropriate for ``local physics'' like quantum entanglement, by contrast with the conventional duality to the eigenvalue density which works well for asymptotic observables like S-matrices. We briefly discuss possible relation of our bosonization to D0 branes.
\end{abstract}

\pagenumbering{roman}

\newpage

\tableofcontents
\pagenumbering{arabic}
\setcounter{page}{1}



\setcounter{footnote}{0} 


\section{Introduction and Summary}\label{sec:intro}

Over the recent decades, spacetime geometry has been described as an emergent notion, in the large $N$ \footnote{Unless otherwise stated, we will denote by $N$ the rank of the matrix.} limit,  from matrix theories, e.g. in the context of AdS/CFT, the BFSS matrix model, the c=1 matrix model, and so on.

A toy model for such emergent notion is the theory of a time-dependent one-matrix model, which is sometimes called matrix quantum mechanics, defined by \eq{matrix-QM} (see also Appendix \ref{app:matrix-QM}). It has been traditionally thought that a good continuum description for large $N$ matrix QM  is provided by the eigenvalue density (also called the “collective field” \cite{Das:1990kaa}). However, there are difficulties with this description since the eigenvalue density possesses infinitely more degrees of freedom compared with the matrix except in the strict $N=\infty$ limit (see Section \ref{sec:prob} for more details). 

In this note\footnote{A preliminary version of this work was presented by one of us in meetings entitled ``Large-N Matrix Models and Emergent Geometry'' (September 4-8, 2023, ESI, Vienna, Austria) and ``Aspects of CFTs'' (January 8-11, 2024, IIT Kanpur, India).}, we will be mostly concerned with a continuum understanding of an {\it entanglement entropy (EE) formula} that arises from matrix quantum mechanics in the large N limit.  

Let us first try to understand what EE means in the context of a matrix model which does not have space. This question has been addressed by using the tool of “target space EE” (TSEE) \cite{Das:2020jhy, Das:2020xoa, Mazenc:2019ety}, specifically for D0 brane models. In case of one-matrix QM, as is well-known (see, e.g. the reviews in \cite{Ginsparg:1993is, Klebanov:1991qa, Polchinski:1994mb}; see also the new interpretations of the $c=1$ model in, e.g. \cite{McGreevy:2003kb, Klebanov:2003km, Douglas:2003up}),  the matrix model in its singlet sector becomes equivalent to a theory of $N$ free fermions moving on the real line\footnote{The $N$ real eigenvalues become position coordinates of $N$ fermions.}, which can be described in terms of a second quantized field, $\psi(\lambda)$. It turns out \cite{Das:2020jhy} that the EE defined in terms of the TSEE can be identified with the normal base space EE of the fermionic QFT.

Many of our results will apply to matrix QM in general, which we will define in terms of an action\footnote{In this paper, we will focus only on the singlet sector of the matrix model.  Throughout the paper, the statement of the equivalence between matrix QM and free fermions is made in this context. In a gauged variant of the theory, e.g. in case of the $c=1$ matrix model in its rolling tachyon interpretation \cite{McGreevy:2003kb, Klebanov:2003km, Douglas:2003up}, the singlet condition is automatically enforced by the presence of a non-dynamical gauge field $A_0(t)$. \label{ftnt:singlet}} (see Appendix \ref{app:matrix-QM} for more details)
\begin{align}
  S= \beta N \int dt \{\frac12 \Tr (\dot M)^2 - \Tr V(M) \}
  \label{matrix-QM}
\end{align}
Here $M(t)$ is a time-dependent Hermitian matrix, $\beta$ is a positive real number. As mentioned above, the eigenvalues of this matrix, $\l_i(t)$ behave like fermions, and each of them is described by a single-particle hamiltonian $\tilde h= \beta N h$, where
\begin{align}
  h =  -\frac1{(\beta N)^2} \frac{\del^2}{\del \l^2} + V(\l)
  = -\hbar^2 \frac{\del^2}{\del \l^2} + V(\l), \; \hbar \equiv \frac1{(\beta N)}
  \label{single-ham}
\end{align}
In the above we have formally identified $\frac1{(\beta N)}$ with an effective $\hbar$. The large $N$ limit is defined by
\begin{align}
  \beta= {\rm fixed}, N \to \infty
  \label{large-N}
\end{align}
A simultaneous tuning of $\beta$, as in \eq{double-scaling}, defines the so-called ``double scaling'' limit which we will discuss shortly.

The EE of the model \eq{single-ham} has been discussed extensively. A large $N$ result, obtained from a Thomas-Fermi treatment of the equivalent fermionic description \cite{Smith:2020gfl, Das:2022mtb} (see also references therein, as well as Appendix \ref{app:thomas-fermi}), goes as follows. Suppose that the matrix model is in its ground state, which is equivalently described by the Slater determinant of the $N$ lowest energy fermion wavefunctions, or, more simply, by the filled Fermi sea. In this state, the EE for an interval $A= (\lambda_1, \lambda_2)\in {\bf R}$ is given by (see Appendix \ref{app:thomas-fermi})
\begin{equation} \label{SS}
S_A = \frac{1}{3}\left(\log\left(\frac{2 P_F(\lambda_0) (\lambda_2 - \lambda_1)}{\hbar}\right) + 1 +\gamma_E \right),
\end{equation}
where $P_F(\lambda_0) = \sqrt{2(E - V(\lambda_0)}$ is the Fermi momentum at some point $\lambda_0$ inside the region $A$; here it is assumed that the Fermi momentum varies so slowly inside the interval $A$, the choice of the precise point $\lambda_0$ within the interval is immaterial.

The remarkable thing about the above formula is the following. Suppose we consider, for example, the case of a {\it hard-box potential} for the fermions\footnote{The hard-box potential can be defined by \eq{matrix-QM} with $V(M)$ defined by the potential function in \eq{hard-box-app} in the limit $\bar V \to \infty$. Alternatively, in the singlet sector, we will define the model in terms of the equivalent fermions which are confined in a box $[-L/2,L/2]$, in the spirit of the hard walls introduced in \cite{Moore:1991sf}.}, which are confined in an interval $[0,L]$. Eq. \eq{SS} in this case becomes \footnote{Note that Bohr-Sommerfeld quantization translates to $\frac{P_F L}{\pi\hbar}  =  N$.\label{ftnt:hbar}} 
\begin{equation} \label{SS-v0}
S_A = \frac{1}{3}\log\left(\frac{N (\lambda_2 - \lambda_1)}{L}\right),
\end{equation}
omitting numerical constants which will not be important for our purpose. Now, if we consider free relativistic bosons in a box of size $L$, at a short distance cut-off $\epsilon=L/M$, then we get the following formula for EE \cite{Holzhey:1994we, Calabrese:2004eu} (see also Appendix \ref{app:herzog}).
\begin{equation} \label{relativistic}
S_A = \frac{1}{3} \log\left(\frac{\lambda_2 - \lambda_1}{\ep}\right)= \frac{1}{3}\log\left(\frac{M(\lambda_2 - \lambda_1)}{L}\right),
\end{equation}
The  similarity between the above two formulae suggests that the ground state EE of the nonrelativistic fermion theory perhaps has a description in terms of a relativistic boson, with
\begin{align}
  M = \alpha N
\label{m=n}  
\end{align}
where $\alpha$ is some purely numerical constant. The existence of a bosonic description by itself is not surprising. After all, we expect a low energy description near the fermi surface in terms of a density wave (see, e.g. \cite{Tomonaga:1950zz, Landau:1980mil}); the description of matrix QM in terms of the eigenvalue density is in keeping with this fact. The surprising thing is that the number of lattice points in the box appears to be given in terms of the fermion number! Equivalently stated, the ultraviolet momentum cut-off in the bosonic description appears to be given by the Fermi momentum (this is visible in both \eq{SS} and \eq{SS-v0}). We will, in fact, find such a bosonic description, which is distinct from the density description (the latter is encapsulated in the collective variable formalism of \cite{Das:1990kaa}, see Section \ref{sec:prob} for some more details). 

\subsubsection*{c=1}

As detailed in Section \ref{sec:c=1}, the $c=1$ matrix model is a special case of matrix QM characterized by a potential with a quadratic maximum, e.g. \eq{double-well}. In this case, in the double scaling limit defined in \eq{double-scaling}, the formula \eq{SS} leads to the following entanglement entropy
\begin{align}
  \label{c=1-EE}
S_A = \frac{1}{3} \log(2(\l_2 - \l_1) \sqrt{2\left(-\mu +\frac{{\lambda}^2}{2}  \right)})
\end{align}
where we have dropped the tilde's from $\tilde \l$ in \eq{c=1-EE-app}. The EE of the $c=1$ model, using the fermionic QFT formalism, was first computed by \cite{Das:1995vj}, and refined in \cite{Hartnoll:2015fca}. The formula given by \cite{Hartnoll:2015fca} is
\begin{align}
  \label{c=1-EE-hartnoll}
	S^{\text{Hartnoll}}_A &= \frac{1}{3}\left( \log\left(\sinh(x_1) \sinh(x_2) \frac{x_2 - x_1}{g_s}\right) + \frac{1}{2}\log\left( \frac{16 x_1 x_2}{(x_1+x_2)^2}\right) + \gamma_E \right)
\end{align}
Here $g_s = \frac1{2\mu}$, and the coordinate $x$ is defined in terms of $\l$ as
\begin{align}
  \label{lam-x}
  \l = \sqrt{2\mu} \cosh(x).
\end{align}
Denoting the interval $A$ by $(x_1, x_2) = (x_0 - l/2, x_0 + l/2)$, and taking the limit $l\ll x_0$, \eq{c=1-EE-hartnoll} reduces to
\begin{align}
  \label{c=1-EE-hartnoll-2}
	S_A &= \frac{1}{3}\left( \log\left( \frac{x_2 - x_1}{g_s}\sinh^2(x_0)\right)\right)
\end{align}
up to terms of order $O(l^2/x_0^2)$ and $O(\exp[-x_0])$. It is easy to see that
this expression agrees with \eq{c=1-EE}, up to an unimportant additive numerical term.

The double scaled $c=1$ matrix model is dual to two-dimensional string theory, with a rough equivalence between the background-subtracted eigenvalue density $\rho(\l,t)$ and the (background-subtracted) closed string tachyon $T(x,t)$ of the 2D string theory, with $\l$ and $x$ defined as above in \eq{lam-x} (see \cite{Das:1990kaa, Sengupta:1990bt, Polchinski:1991uq}, the reviews \cite{Ginsparg:1993is, Klebanov:1991qa}, also the more recent interpretations in \cite{McGreevy:2003kb, Klebanov:2003km}). The coordinate space of the 2D string is $x$, which is measured in units of the string length. Reinserting $l_s$ in the above formula \eq{c=1-EE-hartnoll-2}, we get
\begin{align}
  \label{c=1-EE-final}
	S_A &= \frac{1}{3}\left( \log\left( \frac{x_2 - x_1}{l_s g_s}\sinh^2(x_0/l_s)\right)\right)
\end{align}
This formula has a surprise, as the role of the short distance cut-off seems to be played by
\begin{align}
  \ep= l_s g_s,
  \label{ls-gs}
\end{align}
rather than just the string length $l_s$, which is what one {\it a priori} expects in a string theory. We will find that the EE can be explained in terms of the new bosonic variable which has a natural short distance cut-off equal to \eq{ls-gs}.

\subsubsection*{Summary of the main points}

\begin{itemize}

\item
In Section \ref{sec:exact}, we discuss an exact operator bosonization of non-relativistic fermions in one dimension (equivalently, singlet sector of MQM), that was introduced in \cite{Dhar:2005fg}. The algebra of bosonic operators, generated by a finite number $N$ of creation and annihilation operators, is shown to be exactly equivalent to the fixed-$N$ fermion algebra (see \eq{fermionize1}, \eq{fermionize2} and \eq{bosonizeapp}).

\item
\underbar{\textbf{\textit{The main result}}}\\
Using the above bosonic oscillators, in Section \ref{sec:real-space} we postulate a bosonic field defined on a lattice with a finite number $N$ of points (equivalently a finite lattice spacing). For fermions in a hard box potential of length $L$, the bosonic lattice is taken to be of length $L$ with Dirichlet boundary conditions, with a finite lattice spacing $\ep=L/N$. The EE of a ``region'' of the lattice, consisting of lattice points $j_1, j_1+1, ..., j_2$ at locations $x_1= j_1 \ep, ..., x_2 = j_2 \ep$, reproduces the EE \eq{SS-v0} of the fermion problem for a region $[x_1, x_2]$. For MQM with non-trivial potentials, we show (see Section \ref{sec:non-uniform-lat}) that the bosonic theory needs to be defined on a lattice obtained from the previous case by a coordinate transformation. The EE of the bosonic theory of a region of the lattice again agrees with the corresponding EE \eq{SS} of the fermionic theory. Note that the finite EE of the MQM follows in the bosonic representation from the fact that the bosonic theory is on a lattice with a built-in finite lattice spacing, with a finite number $N$ of lattice points (see Section \ref{sec:finite}). In case of the double-scaled theory of $c=1$, discussed in Section \ref{sec:c=1}, even though $N$ is taken to be infinite, the fermionic EE continues to remain finite, because $N$ cancels out due to the simultaneous scaling of the Fermi level as well as the coordinate, as in \eq{c=1-EE-app}. Since the bosonic lattice EE reproduces the fermionic EE, the limiting procedure is the same, which ensures that the lattice boson reproduces the correct, finite, final formula \eq{c=1-EE-final} for the $c=1$ EE. Note that in the double scaled theory, even though $N=\infty$, the effective lattice spacing transmutes to $\ep= g_s l_s$ which is finite.

\item {\it Integrability and universality of the EE}\\
It is explained in Section \ref{sec:integrable} that irrespective of the matrix model potential $V$, the bosonic Hamiltonian, even though it is interacting, defines an integrable system, with all $N$ occupation numbers conserved. A particular consequence is the universality of the ground state $|0\ran$ (with zero occupation numbers), leading to a universal ground state EE \eq{calabrese-cardy-3}.

\item
In Section \ref{sec:prob} we explain some issues with the collective variable description of matrix QM. In Section \ref{sec:resolution} we show how these are resolved in the exact lattice bosonization introduced in this paper. Besides the issue of the EE, we show in Section \ref{sec:moments} that the moment \eq{i4-SHO} is naturally divergent in the collective theory, whereas it is exactly reproduced in the exact boson theory.

\item
In Section \ref{sec:conclusion} we make some remarks on a possible interpretation of the new bosonization of the $c=1$ model. We collect some of the important details in the Appendices.

\end{itemize}

\section{The exact boson}\label{sec:exact}

In this paper, we will explore an exact bosonization of matrix QM in dealing with the above limitations. The basic framework was laid out in \cite{Dhar:2005fg, Dhar:2006ru, Dhar:2005su}, based on earlier work in \cite{Balasubramanian:2001nh, Suryanarayana:2004ig}. Much of this section will be a review of these works.

\subsubsection*{The giant graviton story}

The exact bosonization has its origin in the story of giant gravitons \cite{McGreevy:2000cw}, which are half-BPS D3 branes wrapped on a 3-sphere fibre of the $S^5$ factor of AdS$_5 \times S^5$. These behave like point particles in the remaining two angular directions $\theta, \phi$, and move on a constant $\theta$ orbit with unit angular velocity $d\phi/dt$. The angular momentum is given by the size of the orbit and is maximized, with a value $N$, when the orbit is the equator $(\theta=\pi/2)$.  The half-BPS condition equates the angular momentum to energy; therefore the giant gravitons are objects with a built-in high energy cut-off, namely $N$. The boundary theory of the giant gravitons is given by the half-BPS sector of ${\cal N}=4$ super Yang-Mills theory on $S^3$ which is entirely characterized by a complex matrix $Z(t)= \Phi_5(t) + i \Phi_6(t)$ which is an adjoint scalar carrying a $U(1)$ R-charge. The upper cut-off on the energy also translates to a maximum R-charge equal to $N$.

The quantum mechanics of the holomorphic matrix $Z(t)$ can be reduced essentially to its (complex) eigenvalues, which, like in case of the hermitian matrix model, behave like fermions \cite{Takayama:2005yq}. The states of the $N$-fermion system are given by the filled levels $f_i, i=1,2,...N$ of single particle harmonic oscillator states, with $0\le f_1 < f_2 < ... < f_N < \infty$. It has been observed in this context \cite{Berenstein:2004kk, Suryanarayana:2004ig} and elsewhere that any set of fermion occupancies $\vec f$ can be mapped to a Young Tableaux by the rule (see Figure \ref{fig:young}) that the number of columns $r_{N-k}$ with $N-k$ boxes is given by the ``gap'' below the $k+1$-th fermion, i.e.
\begin{align}
  r_{N-k}= f_{k+1} - f_k -1, \, k=1,...,N-1, \; r_N=f_1
  \label{rk-fn}
\end{align}

\begin{figure}[]
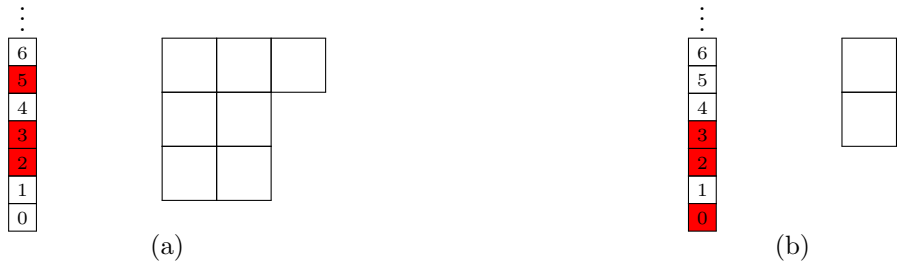

	\begin{minipage}{.5 \linewidth}
		\centering 
\ytableausetup
 {mathmode, boxframe=normal, boxsize=1em}
\begin{ytableau}
  \none[\vdots] \\
  \scriptstyle 6 \\
  *(red) \scriptstyle 5\\
  \scriptstyle 4\\
  *(red)\scriptstyle 3\\
  *(red)\scriptstyle 2\\
  \scriptstyle 1\\
  \scriptstyle 0
  \end{ytableau}
\kern40pt
\ytableausetup
 {mathmode, boxframe=normal, boxsize=2em}
 \begin{ytableau}
   \none\\
 {} & {} & {} \\
 {} & {} \\
 {} & {} 
 \end{ytableau}
 \centerline{(a)}
  	\end{minipage} \hfill
	\begin{minipage}{0.5 \linewidth}
		\centering 
    \ytableausetup
 {mathmode, boxframe=normal, boxsize=1em}
\begin{ytableau}
  \none[\vdots] \\
  \scriptstyle 6 \\
  \scriptstyle 5\\
  \scriptstyle 4\\
  *(red)\scriptstyle 3\\
  *(red)\scriptstyle 2\\
  \scriptstyle 1\\
  *(red) \scriptstyle 0
\end{ytableau}
\kern40pt
\ytableausetup
 {mathmode, boxframe=normal, boxsize=2em}
 \begin{ytableau}
   \none\\
 {} \\
 {}  
 \end{ytableau}
 \centerline{(b)}
	\end{minipage}
	\caption{\footnotesize (a): Here the left column denotes the fermion filling: there are $N=3$ fermions occupying the levels 2,3,5, i.e. $\vec f= \{2,3,5\}$. The gaps below fermion numbers 1,2 and 3 (counting from below) are, respectively, 2,0,1. This corresponds to the Young Tableau on the right: with $(r_3, r_2, r_1)=(2,0,1)$, i.e. there are 2 columns with 3 boxes each, 0 columns with 2 boxes and 1 column with 1 box. (b): Here the filled levels are $\vec f= \{0,2,3\}$, yielding $(r_3, r_2, r_1)=(0,1,0)$ which describes a single column YT of height 2, corresponding to an antisymmetric 2-tensor representation. In both (a) and (b) the precise correspondence between the YT and the filled fermion levels is that the Schur polynomial for the representation $R$ characterizing YT, acting on the filled Fermi sea, produces an excited $N$-fermion state characterized by the filled fermion levels depicted.}
	\label{fig:young}
\end{figure}
The map \eq{rk-fn} is not a mere formality. A multi-giant graviton state with $r_k$ giant gravitons carrying angular momentum $k$, $k=1, 2, ...,N$, is created by an operator $\chi_R(Z)$ acting on the ground state of the system, where $\chi_R(Z)$ is a Schur polynomial for a representation $R$ characterized by a Young tableau with $r_k$ columns each with $k$ boxes, $k=1, 2, ...,N$ \cite{Balasubramanian:2001nh, Corley:2001zk, Berenstein:2004kk, Suryanarayana:2004ig}. It can be shown that such a state indeed has filled fermion levels $f_n$ related to the $r_k$'s by the map \eq{rk-fn}. See Appendix \ref{app:schur} for further details.

\subsection{Exact operator bosonization}\label{sec:exact-op}

While \eq{rk-fn} gave a map between the giant graviton states and the fermion states, and it was found that the Schur operators characterized by $\{r_k\}$, acting on the ground state, created a fermion state given by $\{f_n\}$, it was found that the action by successive Schur operators on the ground state produced a mixture of Slater determinants which were not characterized by a single set of filled fermion levels. It was therefore not immediately obvious what operators took us from one set of filled Fermi levels to another set of filled Fermi levels. This problem was somewhat nontrivial, and was solved in \cite{Dhar:2005fg} and further developed in \cite{Dhar:2006ru, Dhar:2005su}. We will describe the results below briefly. We will present it as an exact operator bosonization of nonrelativistic fermions in a 1D confining potential. The connection to the giant graviton story is not difficult to work out. The connection to general matrix QM (with a hermitian matrix $M(t)$) is obvious.

We consider a free, non-relativistic $N$-fermion system in a confining potential in one dimension. The single particle energy eigenfunctions are assumed to be $\chi_i(\l)$ with discrete, non-degenerate energy values $\vep_i$, $i=0,1,2,.. \infty$:
\begin{align}
  h(\l, \del_\l) \chi_i(\l) = \vep_i \chi_i(\l)
  \label{single-spectrum}
\end{align}
A simple basis of the $N$-particle Hilbert space is provided by the Slater determinants $\Psi(\l_1, ..., \l_N)= {\rm Det}_{ij} \chi_{f_i}(\l_j)$ which represents states $| \vec f\ran$ with filled fermion levels $f_i, i=1,2,...,N$. It is useful to introduce a second-quantized Fermion field
\begin{align}
  \label{second-q}
  &\Psi(\l)= \sum_{i=0}^\infty \chi_i(\l) \psi_i, \\
  &\int d\l \ \Psi^\dagger(\l) \Psi(\l)=N
\end{align}
in terms of which
\begin{align}
  \label{f-ket}
  | \vec f\ran =   \psi_{f_N}^\dagger ... \psi_{f_2}^\dagger\ \psi_{f_1}^\dagger |0\ran_F
\end{align}
The second line of \eq{second-q} ensures that we are working with a fixed fermion number; the allowed operator algebra is generated by bilinears of the type $\psi_{k}^\dagger \psi_l$.

The operator bosonization, which we found in \cite{Dhar:2005fg} and will use below, maps these states to the following bosonic states:
\begin{align}
  \label{identity}
  | \vec f \ran = | \vec r \ran \equiv \prod_{n=1}^N \frac{(a_n^\dagger)^{r_n}}{\sqrt{r_n !}}|0\ran_B
\end{align}
where $r_k$ and $f_k$ are related by \eqref{rk-fn}.
In this bosonization {\it\underbar{a finite number $N$}} of oscillators $a_n, a_n^\dagger$, $n=1, 2, ..., N$ is used, which satisfy an exact Heisenberg algebra
\begin{align}
  \label{heisen}
        [a_n, a_m^\dagger ]= \delta_{nm}, \; n,m=1,2,...,N
\end{align}
The $|0\ran_B$ is the no-particle state in the bosonic theory, annihilated by all the $a_n$. Note that we have presented \eq{identity} as an equality, in the sense that the LHS and RHS are bosonic and fermionic representations of the {\it\ub{same state}}.

The basic ingredients of the construction, very briefly, involve introducing ``raising'' operators $\sigma^\dagger_k$ (see Figure \ref{fig:sigma-k})  and their adjoints.
\begin{figure}
\begin{center}
\includegraphics[scale=.5]{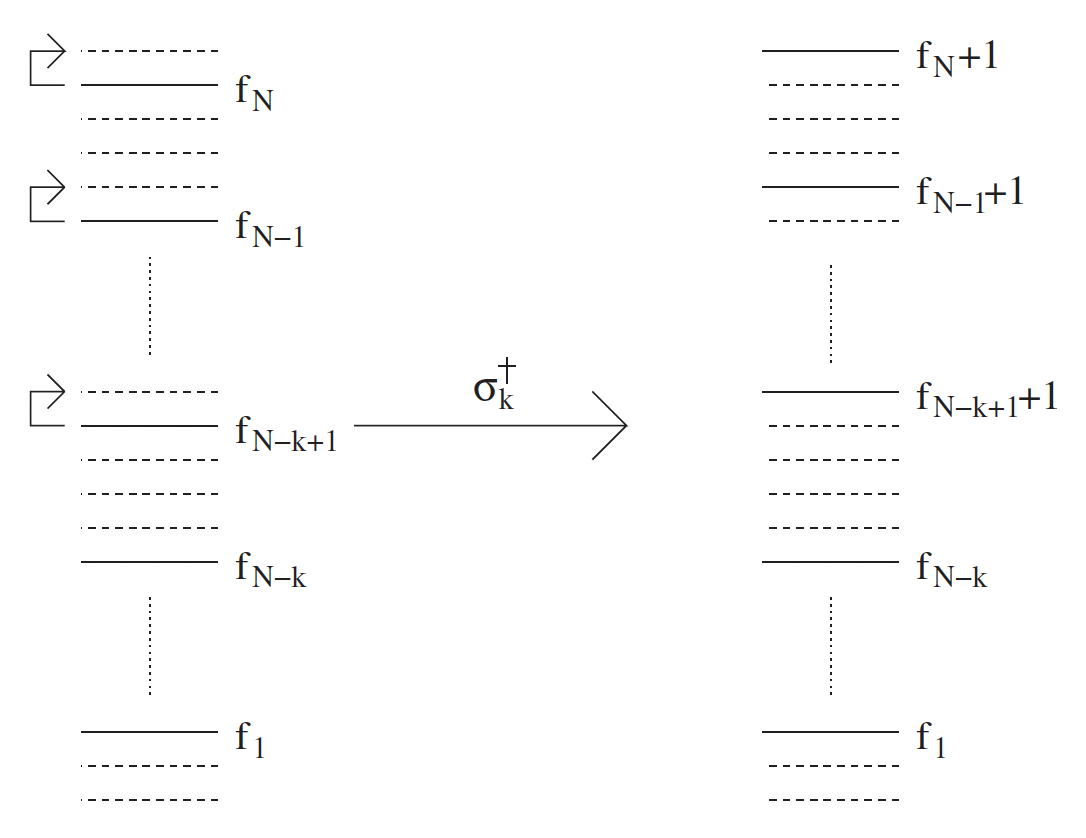}
\end{center}
\caption{\footnotesize The action of $\sigma_k^\dagger$ on the state $|\vec f \ran$ pushes the top $k$ fermions up by one level, starting from top down. Clearly, if we act on the filled Fermi sea with $\vec f={0,1,2,...,N-1}$, one gets the state $\vec f={0,1,2,.., N-k-1, N-k+1,N-k+3, ...,N}$, i.e. the operator creates a hole at depth $k$ (see also Figure \ref{fig:young}; also compare with the action of the Schur of antisymmetric $n$-tensor as detailed in Appendix \ref{app:schur}). The action of the adjoint $\sigma_k$ involves pushing the top $k$ fermions down, starting from down up, with the understanding that if the target level is occupied, it annihilates the state. The picture is taken from \cite{Dhar:2006ru}.}
\label{fig:sigma-k}
\end{figure}
From the action of $\sigma_k, \sigma_k^\dagger$ it follows that, on a state $| \vec f \ran$, 
\begin{align}
  \label{sigma-alg}
  \sigma_k \sigma_k^\dagger =1,\; \sigma_k^\dagger \sigma_k = \theta(r_k-1), \;
        [\sigma_k, \sigma_l^\dagger]=0 \hbox{ if } k\ne l
\end{align}
where $r_k \equiv f_{N-k+1}- f_{N-k} -1 $  and $\theta(m) \equiv 1$ if $m\ge 0$.

The bosonic oscillators satisfying \eq{heisen} are related to these $\sigma$-operators by
\begin{align}
  \sigma_k= \frac1{\sqrt{a_k^\dagger a_k +1}} a_k, \;
  \sigma_k^\dagger = a_k^\dagger \frac1{\sqrt{a_k^\dagger a_k +1}}
  \label{sig-a}
\end{align}

Note that \eq{identity} implies that the bosonic ground state coincides with the Fermion ground state. To see this, put $r_n =0$ in the RHS of \eq{identity} for all $n$; by the map \eq{rk-fn}, we find that
$\vec f = \vec f_0 \equiv \{0,1,2,...,N-1\}$ i.e. the filling corresponding to the filled Fermi sea; hence
\begin{align}
  |0\ran_B =|F_0 \ran \equiv | \vec f_0 \ran, 
  \label{ground-ground}
\end{align}

\subsubsection{The bosonic Hamiltonian}

Suppose the fermionic theory has a single-particle Hamiltonian
$\hat h= p^2/(2m) + V(\l)$ with the following spectrum:\footnote{From here on, we will put $m=1$ for simplicity; it can always be put back to correct the dimensionality of various formulae.\label{ftnt:m=1}}
\begin{align}
\hat h |n\ran= \vep(n) | n \ran\ \leftrightarrow \ (-\fr{\hbar^2}2 \del_\l^2 + V(\l)) \chi_n(\l) = \vep(n) \chi_n(\l), \;
n=0,1,...,\infty,
\label{spectrum}
\end{align}
where we assume a non-degenerate spectrum with
\begin{align}
  \vep(0) < \vep(1) < ...
  \label{ordering}
\end{align}

Thus, for a particle in a hard box of length $L$, where $\hat h= p^2/2$ with vanishing boundary condition for the wavefunction at $x=0, L$, we have\footnote{Note the appearance of $n+1$ in the spectrum; this is because we want to label the mode numbers starting from $n=0$.}
\begin{align}
  \chi_n(\l)= \sqrt{\fr2{L}} \sin(\fr{(n+1)\pi \l}{L}), \;
  \vep(n)=\fr\a2 (n+1)^2,\; \a= \fr{\hbar^2 \pi^2}{L^2}, \quad n=0,1,2,...,\infty
\label{hard-box}
\end{align}
A fermionic state $|F \ran \equiv |\vec f \ran$, described by a filling
$\vec f=\{ f_1, f_2, ..., f_N \}$, clearly has energy
\begin{align}
  E= \sum_{n=1}^N \vep(f_n)
  \label{fermi-energy}
\end{align}
The question is, what bosonic Hamiltonian, acting on the corresponding bosonic state $|\vec r \ran$, has the same energy? To proceed, let us first find the inverse of the relations \eq{rk-fn}:
\begin{align}
  f_1= r_N,
  ...,\ f_{N-k+1} = r_N + ... + r_{k}  +(N-k), ...,
  f_N= r_N + ... + r_1  + (N-1)
  \label{fn-ek}
\end{align}
Substituting these in \eq{fermi-energy} we get
\begin{align}
  E=\sum_{n=1}^N \vep(f_n)= \sum_{k=1}^{N} \vep(f_{N-k+1}) = \sum_{k=1}^{N} \vep(\sum_{i=k}^N r_{i}  +(N-k))
  \label{bose-energy}
\end{align}
The desired bosonic Hamiltonian is, clearly, then
\begin{align}
  H_{bose}= \sum_{k=1}^{N} \vep\left(\sum_{i=k}^N a_i^\dagger a_i  +(N-k)\right)
  \label{bose-ham}
\end{align}
since, $a_i^\dagger a_i$, acting on $|\vec r \ran$, gives $r_i$.

\paragraph{Exact ground state}
Note that, because of the monotonicity condition \eq{ordering}, $H_{bose}$ is minimized when $a_i^\dagger a_i=0$; hence the state $|0\ran_B$, defined by
$a_k |0\ran_B =0$, remains the exact ground state of the (possibly very complicated and interacting) Hamiltonian $H_{bose}$, with ground state energy
\begin{align}
  E_{bose, ground}= \sum_{k=1}^{N} \vep(N-k)= \sum_{k=0}^{N-1} \ep(k) = E_{fermi, ground}
  \label{bose-ground}
\end{align}

\subsubsection*{Hard box}
For the hard box spectrum \eq{hard-box}, the bosonic Hamiltonian \eq{bose-ham} becomes
\begin{align}
 &\kern-25pt H_{bose}=\fr{\hbar^2 \pi^2}{2 L^2} \sum_{k=1}^{N} \left(\sum_{i=k}^N a_i^\dagger a_i + N-k+1 \right)^{\!{}_2}
  \label{bose-ham-hard-box}
\end{align}
which can be expanded as
\begin{align}
  & H_{bose} = E_g + H_1 + H_2 \label{bose-expand}\\
  & H_1 = \a N \sum_{k=1}^N k a^\dagger_k a_k
  \label{h-1}\\
  & H_2 = \fr\a{2} \sum_{k=1}^N \left( - k(k-1) \sum_{i=k}^N a^\dagger_i a_i 
  + \left(\sum_{i=k}^N a^\dagger_i a_i \right)^{\!{}_2} \right)
  \label{h-2}
\end{align}
Here $E_g$ is the ground state energy \eq{bose-ground} evaluated for the hard box, and is given by
\begin{align}
  E_g = \fr\a{12} (2N^3 + 3N^2 + N), \; \a= \fr{(\hbar\pi)^2}{L^2}
  \label{e-g}
\end{align}  
The energy above the ground state is given by $H_1 + H_2$. At low energies, i.e. for $O(1)$ occupation numbers for $r_k$'s with small $k$, $H_1 = O(N)$, $H_2 = O(1)$.  In Appendix \ref{app:lattice-partn-fn}, we show that at temperatures lower than $O(N)$ (which is not much of a restriction at large $N$) $H_1$ is the dominant contribution.

It is interesting to note that the linear part of the bosonic Hamiltonian $H_1$ can be written as \footnote{Note that we have $v_F=p_F$ because of the $m=1$ convention, see footnote \ref{ftnt:m=1}.}
\begin{align}
  & H_1 = \sum_{n=1}^N \ep_n a^\dagger_n a_n, \hbox{where} \nonumber\\
  & \ep_n = N \a n = v_F p_n, \;\; v_F=p_F= \fr{\hbar \pi N}{L},\, p_n=\fr{\hbar \pi n}{L} \nonumber\\
  & {\hat\omega}_n \equiv \fr{\ep_n}\hbar = v_F k_n, \;\; k_n= \fr{p_n}\hbar = \fr{\pi n}{L}
  \label{fermi-level-dispersion}
\end{align}
which corresponds with the single-particle fermionic dispersion near the Fermi surface
\begin{align}
\ep(p) \equiv \fr{(p_F + p)^2}{2m} - \fr{p_F^2}{2m} \approx v_F p,
\label{fermi-vel}
\end{align}
where $v_F= p_F/m$ is the Fermi velocity (in this paper we have used the convention $m=1$). In the above, we have considered the hard box boundary conditions. In Appendix \ref{app:periodic}, we have discussed fermions in a periodic box and its corresponding exact bosonization.

\subsubsection*{Harmonic Oscillator}
The exact boson Hamiltonian for the case of fermions in a simple harmonic oscillator potential of frequency $\omega$ takes on the following simple expression that is linear in the occupation number operators. \footnote{A bosonization of finite $N$ matrix quantum mechanics in the simple harmonic oscillator potential is carried out in \cite{Itzhaki:2004te}, which at first glance may seem similar to ours, especially with their bosonic Hamiltonian matching with ours. However, in their bosonization, the bosonic mode cut-off is put in by hand while our bosonic oscillators have a natural cut-off equal to $N$.}
\begin{align}
	H_{bose} &= \hbar \omega \sum_{k=1}^N \left(\sum_{i=k}^N a_i^{\dagger} a_i + N - k + \frac{1}{2}\right) \nonumber \\
	&= E_g + \hbar \omega \sum_{k=1}^N k a_k^{\dagger} a_k 
\end{align}
As before, $E_g$ is the ground state energy 
\begin{align}
E_g =\frac{1}{2} \hbar \omega N^2
\end{align}

\subsubsection*{More general potential}
The bosonic Hamiltonian in the more general case also has a low energy expansion, similar to the $E_g + H_1$ expansion in the hard box case. In the low energy approximation, $r_k$ is $O(1)$ for small $k$ and zero for all higher $k$. The summation over $k$ in the Hamiltonian can be investigated separately in the small $k$ sector and the large $k$ sector as follows. 

\begin{align}
H &= \sum_{k=1}^N \epsilon\left(\sum_{i=k}^N r_i + N - k\right) \nonumber \\
&= \left(\sum_{\text{small $k$}} + \sum_{\text{large $k$}}\right) \epsilon\left(\sum_{i=k}^N r_i + N-k\right) \label{split-general-hamil}
\end{align} 
The Bohr Sommerfeld quantization formula 
\begin{align}
\int_{a_n}^{b_n} \sqrt{2(E_n - V(x"))} dx' = n \pi \hbar
\end{align}
can be used to obtain energy differences near the Fermi level as shown below
\begin{align}
n \pi \hbar &= \int_{a_{N+n}}^{b_{N+n}} \sqrt{2(E_{N+n} - V(x'))} dx' - \int_{a_N}^{b_N} \sqrt{2(E_N - V(x"))} dx' \nonumber \\
&\approx  (E_{N+n} - E_N) \int_{a_N}^{b_N} \frac{1}{\sqrt{2(E_N - V(x'))}} dx' \nonumber \\
&\approx (E_{N+n} - E_N) \frac{\tau_N}{2}
\end{align}
Using the above formula, in the small $k$ sector, we can write 
\begin{align} 
\epsilon\left(\sum_{i=k}^N r_i + N -k \right) &\approx \epsilon(N-k) + \frac{2 \hbar \pi}{\tau_{N-k}}  \sum_{i=k}^N r_i \nonumber \\
 &\approx  \epsilon(N-k) + \frac{2 \hbar \pi}{\tau_N}  \sum_{i=k}^N r_i \label{small-k}
\end{align}
And in the large $k$ sector, since $r_i$ for $i>k$ are zero, we effectively obtain
\begin{align} 
\epsilon\left( \sum_{i=k}^N r_i + N-k \right) &\approx \epsilon(N-k) \nonumber \\  &\approx  \epsilon(N-k) + \frac{2 \hbar \pi}{\tau_N}  \sum_{i=k}^N r_i  \label{large-k}
\end{align} 
Plugging \eqref{small-k} and \eqref{large-k} in \eqref{split-general-hamil} (and calling $2 \pi / \tau_N$ as $\omega_0$) yields
\begin{align}
H &= \sum_{k=1}^N \left( \epsilon(N-k) + \hbar \omega_0 \sum_{i=k}^N r_i \right) \nonumber \\
&= \sum_{k=1}^N \epsilon(N-k) + \hbar \omega_0  \sum_{k=1}^N k r_k  \label{omega0}
\end{align}
where the first term is the ground state energy $E_g$ and the second term $\hbar \omega_0 \sum_k k a^{\dagger}_k a_k$ is the dominant contribution in the low energy approximation. One can compute $\hbar \omega_0$ for the hard box potential and see that it indeed equals $\alpha N$. 

\subsection{General properties of the exact boson}\label{sec:gen-prop}

\subsubsection{Comments on operator algebra}

In the fermionic language, the complete operator algebra of the fixed fermion number theory is clearly generated by the set of fermion bilinears of the form $\Phi_{mn}=\psid_m \psi_n$. The commutation algebra of these bilinears is of the form
\begin{align}
  \label{w-alg}
  [\Phi_{mn}, \Phi_{rs}]= \delta_{nr} \Phi_{ms} - \delta_{ms} \Phi_{nr}
\end{align}
In the context of the $c=1$ matrix model this algebra has been studied in detail and has been identified with the $W_\infty$ algebra \cite{Dhar:1992hr, Dhar:1992rs, Das:1991uta}. More generally, for one dimensional non-relativistic fermi systems (equivalent to the matrix quantum mechanics \eq{matrix-QM}), $\Phi_{mn}$'s play a dual role: (a) they are linearly related to the second quantized Wigner phase space distribution operator, which, in the large $N$ limit, describes, $N$-fermion states in terms of droplets in phase space, (b) they transform one state to another by the action of $W_\infty$ \cite{Dhar:1992hr, Dhar:1992rs, Mandal:2013id, Kulkarni:2018ahv}.

As discussed in Section \ref{sec:prob}, the ``collective'' variables $\rho(\l)$ and $\Pi(\l)$ are linear combination of these fermion bilinears and form part of the $W_\infty$ algebra, and do not have a closed commutation algebra, e.g. the Heisenberg algebra, by themselves. 

The case of the bosonic oscillators $a_k, \ad_l$ found above, however, is different. They satisfy the precise Heisenberg commutation \eq{heisen}, and by construction, are expressible in terms of fermion bilinears (since the $\sigma$-operators involve pushing individual fermion levels up or down). The precise relation, worked out in \cite{Dhar:2005fg}, is 
\begin{align}
\ad_k \equiv & \sum_{m_k> m_{k-1} > \cdots > m_0}
\sqrt{m_1 - m_0}~(\psid_{m_0}\psi_{m_0})(\psid_{m_1 + 1} \psi_{m_1}) \cdots
(\psid_{m_k + 1} \psi_{m_k}) \nonumber \\
& ~~\times \delta \biggl (\sum_{m=m_0 +1}^{m_1 -1} 
\psid_m \psi_m \biggr )~
\delta \biggl (\sum_{m=m_1 +1}^{m_2 -1}\psid_m \psi_m \biggr ) \cdots
~\delta \biggl (\sum_{m=m_{k-1} +1}^{m_k -1} \psid_m \psi_m \biggr ) \nonumber \\
& ~~\times 
\delta \biggl (\sum_{m=m_k +1}^{\infty}\psid_m \psi_m \biggr ), 
\quad \quad \quad \quad k=1, 2, \cdots , (N-1)
\label{fermionize1}
\end{align}
and
\begin{align}
\ad_N \equiv & \sum_{m_N> m_{N-1} > \cdots > m_1}
\sqrt{m_1 + 1}~(\psid_{m_1 + 1} \psi_{m_1}) \cdots 
(\psid_{m_N + 1} \psi_{m_N}) \nonumber \\
& ~~~~~~~~~~~~\times 
\delta \biggl (\sum_{m=m_1 +1}^{m_2 -1}\psid_m \psi_m \biggr ) \cdots
~\delta \biggl (\sum_{m=m_{N-1} +1}^{m_N -1} \psid_m \psi_m \biggr ) \nonumber \\
& ~~~~~~~~~~~~\times 
\delta \biggl (\sum_{m=m_N +1}^{\infty}\psid_m \psi_m \biggr ).
\label{fermionize2}
\end{align}
The expression for the annihilation operators $a_k$, $k=1,2,...,N$ is given by taking the adjoint of the above equations.

The surprising thing is that the bosonic operators defined by the intricate relations above, together with \eq{w-alg}, lead to the very simple bose commutation relations \eq{heisen}! As we promised, these oscillators, therefore, are exact bosons.

\subsubsection{A fuzzy phase space}\label{sec:fuzzy}

The finite number of oscillators $a_k, a^\dagger_k$, $k=1,2,...,N$ can be interpreted in terms of a finite dimensional single-particle Hilbert space ${\cal H}_1 = {\rm Span}\{|k \ran, k=1,2,...,N\}$, where $|k \ran = \ad_k|0\ran_B$. As mentioned in \cite{Dhar:2005fg}, this corresponds to a compact phase space; in case of harmonic oscillator levels, this can be interpreted as a fuzzy disc (see \cite{Lizzi:2003ru, Pinzul:2001my}). One way to see the compactness is to note that the Husimi distributions in phase space corresponding to the states $|k\ran$ have a maximum radius $r \sim \sqrt N$; the `fuzzy' nature follows from the fact since $\fr{r^2}2$ and $\theta$ are canonically conjugate variables, a finite cut-off on the radius implies that the angular resolution cannot be finer than $\Delta \theta \sim 1/N$. One can also see fuzziness in coordinate space, since the finite dimensional ${\cal H}_1$ does not span $L^2({\bf R})$. 

\gap3

\vbox{
\noindent \begin{minipage}{300pt}
    For multi-particle states, the occupancy of each level can be represented in terms of heights of Husimi distributions of the corresponding orbits on the compact phase space. See \cite{Dhar:2005fg} for more details, from which the figure on the right is reproduced.
  \end{minipage}
\hfill
  \begin{minipage}{150pt}
    \begin{figure}[H]
  \centering
  \includegraphics[scale=.4]{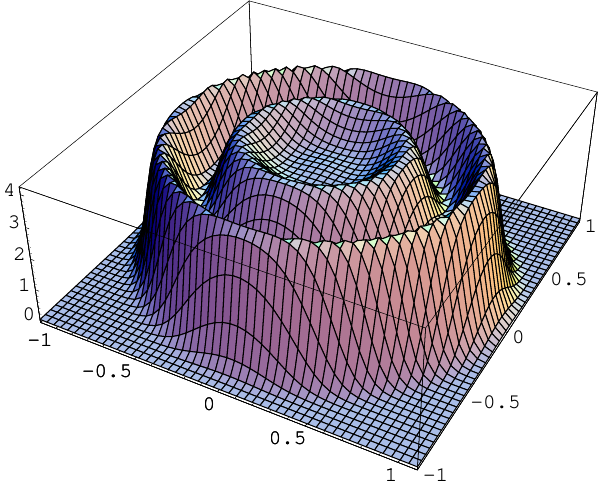}
    \end{figure}
  \end{minipage}
  }

\gap3

\subsubsection{Relation to bosonic representation of $SU(K)$}\label{sec:su-k}

A generalization of the exact bosonization discussed above exists where the single-particle fermion levels are finite in number, say $K$. For $N \le K$, the $N$-particle states correspond to the completely antisymmetric $N$-tensor irrep of $SU(K)$ (corresponding to a single Young Tableau with $N$ boxes). The bosonization corresponds to a bosonic construction of the same representation (see Appendix C of \cite{Dhar:2005fg}).\footnote{A somewhat related approach to bozonization of fermions can be found in the old works \cite{Garbaczewski:1974rp, Garbaczewski:1977fw}. We thank Piotr Garbaczewski for bringing these papers to our attention.}

\subsubsection{The exact boson theory is integrable}\label{sec:integrable}

The bosonic hamiltonian corresponding to the free fermion theory, is typically ``interacting'', by which we mean that a hamiltonian like \eq{bose-expand}, is non-quadratic. More specifically, in terms of the lattice boson we will introduce shortly, \eq{bose-expand} will contain quartic terms in terms of the lattice field. How is it possible that the exact bosonization of a free fermion theory leads to an interacting bosonic theory?

It turns out that the bosonic theory, although it is interacting in the above sense, is actually integrable.\footnote{We thank Shiraz Minwalla for pointing this out to us.} It is perhaps not unexpected that the bosonic Hamiltonian is integrable since it is given solely in terms of the occupation number operators and hence, it is clear that all the $N$ occupation numbers $\hat{N_k}= a^\dagger_k a_k$, $k=1,2,...,N$ are conserved quantities (these are infinite number of commuting conserved charges for $N\to \infty$). The Hamiltonian is \\
(a) diagonal in the occupation number eigenstates $|r_1, r_2, ..., r_N \ran$ \eq{identity}, \\
and (b) completely specified in terms of these numbers $r_i$, as can be inferred from \eq{bose-ham}:
\begin{align}
  H_{bose}= \sum_{k=1}^{N} \vep\left(\sum_{i=k}^N r_i  +(N-k)\right),
  \label{bose-ham-ri}
\end{align}
Just as dynamical transitions are not allowed in the fermionic theory between fermion number eigenstates \eq{f-ket}, similarly dynamical transitions are not allowed in the bosonic theory between various occupation number eigenstates. 

\subsubsection*{GGE}

In case of integrable models, the standard Gibbs ensemble is replaced by the more refined notion of a generalized Gibbs ensemble (GGE, see, e.g. \cite{Rigol:2007a, Calabrese:2012}), defined by the GGE partition function
\begin{align}
  Z_{GGE} & = \tr \left(\exp[-\mu_k \hat{N_k}]\right)
  =\sum_{r_1=0}^\infty... \sum_{r_N=0}^\infty \lan r_1, r_2, ..., r_N | \exp[-\mu_k \hat{N_k}] |r_1, r_2, ..., r_N \ran
  \nonumber\\
  &= \sum_{r_1=0}^\infty... \sum_{r_N=0}^\infty \exp[-\mu_k r_k] = \prod_{k=1}^N (1- \exp[-\mu_k])^{-1}
  \label{GGE-part}
  \end{align}
The ``chemical potentials'' $\mu_k$ (which are sometimes called mode-dependent ``temperatures'' $\beta_k$ \cite{Calabrese:2012}) are given in terms of the average value of $\hat N_k= \bar r_k $, as follows:
\begin{align}
  \bar r_k =\del_{\mu_k} \log Z_{GGE}=  \fr1{e^{\mu_k}-1}
  \label{qk-mu-k}
\end{align}

\section{Constructing a lattice field theory for the exact bosons and computation of EE}\label{sec:real-space}

We saw in the above section that the exact bosons found above represent a fuzzy or granular geometry. In this section, we will define a different realization of the fuzziness, in terms of a lattice. We will find that although the construction of the oscillators in terms of the fermionic oscillators is universal (independent of the fermion potential $V(\l)$), see \eq{fermionize1}, \eq{fermionize2}, the construction of the lattice field $\phi(x)$ in terms of the oscillators, will depend on $V$, much like the construction of the second quantized fermion field \eq{second-q} in terms of the fermionic oscillators depends on $V$. \footnote{The lattice bosonization described in this paper is different from the lattice boson proposed in \cite{Dhar:2005fg}.}

\subsection{Box potential}\label{sec:fixed-lat}

We have seen that corresponding to free fermions in a one-dimensional box, the bosonic Hamiltonian is of the form given by \eq{bose-expand}, which at low energies (or low temperature) becomes $H_1$ \eq{h-1} which has a linear spectrum $\omega(k) = |k|$. It is natural, therefore, to adopt the discussion of lattice version of a relativistic field given in Appendix \ref{app:herzog}. The discussion in Appendix \ref{app:lattice-box} was for a general number $M$ of lattice points, which involved $M$ pairs of oscillators. In the present case, the number of oscillator pairs $a_k, \ad_k$ is fixed to be $N$, the fermion number. Hence, we put $M= N$ (this is as required from \eq{m=n}), and consider a lattice of size $L= (N+1) \ep$. Rewriting \eq{phi-def-hard-app}, \eq{pi-def-hard-app}, with $M \to N$, and evaluating at $t=0$, we get the following definition of the lattice field
\begin{align}
  \phi_j &= \sum_{n=1}^{N} \frac{1}{\sqrt{2 \omega_n}}\sqrt{\frac{2}{L}}\sin(\frac{n \pi  j}{N+1})\left[a_n  + a^{\dagger}_n\right] \label{phi-def-hard} \\
	\pi_j &= \sum_{n=1}^{N} i\epsilon \sqrt{\frac{\omega_n}{2}}\sqrt{\frac{2}{L}}\sin(\frac{n \pi  j}{N+1}) \left[-a_n + a^{\dagger}_n\right], \label{pi-def-hard}
\end{align}
with 
\begin{align}
  \omega_n = v_F \frac{2}{\epsilon}\sin(\frac{n \pi}{2(N+1)}).
  \label{omega-n-box}
\end{align}
Here $v_F$ is the Fermi velocity defined in \eq{fermi-level-dispersion}, \eq{fermi-vel}.

We note that the lattice size $L$ in the bosonic theory is taken to be the same as the box size of the Fermionic theory. For, if the bosonic lattice was of size $\tilde L \ne L$, then the frequency \eq{omega-n-box} would have had Fermi-velocity $\tilde v_F= \hbar \pi N/\tilde L$ which would differ from the the Fermi-velocity $v_F$ in \eq{fermi-level-dispersion}; consequently, the lattice boson Hamiltonian would not match the low energy part $H_1$ of the exact boson Hamiltonian \eq{bose-expand}.

\paragraph{Exact lattice bosonization}
Note that, although we motivated the above definition of the lattice boson by appealing to the low energy part  $H_1$ of the Hamiltonian \eq{bose-ham-hard-box}, the definition is exact, and constitutes {\it an exact lattice bosonization of the Fermion theory.} To see this, note that\\
(a) The lattice variables satisfy the exact equal time Heisenberg commutation relation:
\begin{align}
  [\phi_j, \pi_{j'}]= i \delta_{jj'}
  \label{heisen-lat}
\end{align}
\noindent (b) The lattice boson can be written in terms of the oscillators $a_k, a^\dagger_{k'}$ and vice versa, the inverse relations being  
\begin{align}
  & a_n = \frac{1}{N+1} \frac{L}{2} \sum_{j=1}^{N} \sin(\frac{n \pi j}{N+1}) \left(\sqrt{2 \omega_n} \phi_j + \frac{i}{\epsilon}\sqrt{\frac{2}{\omega_n}} \pi_j\right)
  \nonumber\\
  & \ad_n = \frac{1}{N+1} \frac{L}{2} \sum_{j=1}^{N} \sin(\frac{n \pi j}{N+1}) \left(\sqrt{2 \omega_n} \phi_j - \frac{i}{\epsilon}\sqrt{\frac{2}{\omega_n}} \pi_j\right)
  \label{inverse-a-ad-box}
\end{align}
Using this fact  and the fact that the oscillators $a_k, a^\dagger_{k'}$  are in one-to-one correspondence with fermion bilinears $\psi^\dagger_n \psi_m$ (see \eq{fermionize1},\eq{fermionize2} and \eq{bosonizeapp},  we see that the lattice boson variables $\phi_j, \pi_{j'}$ are also in exact one-to-one correspondence with the fermion bilinears. Thus the operator algebra of the lattice boson theory is isomorphic to the operator algebra of the fermion theory at fixed fermion number $N$.\\
\noindent (c) The Hamiltonian \eq{bose-ham-hard-box}, equivalently \eq{bose-expand}, can be expressed exactly in terms of the lattice by expressing the oscillators in terms of the lattice boson using \eq{inverse-a-ad-box}. The full expression is non-local and not particularly illuminating. But at low energies, the dominant part of the Hamiltonian \eq{bose-expand} is $H_1$ (see comments below \eq{e-g}); further, at low energies (low wave numbers) the dispersion relation \eq{omega-n-box} becomes linear $\omega_n \sim n$, hence $H_1$ reduces to the standard oscillator Hamiltonian (\eq{oscill-ham-app} with $M \to N$), which, therefore boils down to the standard lattice Hamiltonian \eq{lattice-ham} for a massless boson:
\begin{align}
  & H_{lattice}= \frac{1}{2\epsilon} \sum_j\left[  \pi_j^2 + v_F^2 (\phi_{j+1} - \phi_j)^2 \right]
  \label{lattice-ham-text}
\end{align}
In Appendix \ref{app:lattice-partn-fn} it is shown that the partition function resulting from the above lattice Hamiltonian at temperatures lower than $O(N)$ (which is not much of a restriction at large $N$) reproduces that of the fermion Hamiltonian in the box.

Let us re-emphasize that although the form of the local bosonic fields were determined by using low energy arguments, the lattice boson given in \eqref{phi-def-hard} and \eqref{pi-def-hard} is exact and is valid in any regime. The form of the exact boson Hamiltonian in terms of these lattice bosons will in general be very non-local. However, in the low-energy/semi-classical regime, the exact boson Hamiltonian takes a very simple, local form given in \eqref{lattice-ham-text}.

Also note that we could redefine the field $\phi(x)$ and $\pi(x)$ to $\tilde{\phi}(x) = \phi(x) \sqrt{L}$ and $\tilde{\pi}(x) = \pi(x) / \sqrt{L}$ (such a rescaling doesn't affect the algebra) to rewrite everything purely in terms of $\omega_0$, the coefficient of the linear part of the low energy expansion of the exact Hamiltonian in \eqref{omega0}. 
\begin{align}
\tilde{\phi}_j &= \sum_{n=1}^N \frac{1}{\sqrt{\omega_n}} \sin(\frac{n \pi j}{N+1}) [a_n + a^{\dagger}_n ]  \\
\tilde{\pi}_j &= \sum_{n=1}^N \frac{i}{N} \sqrt{\omega_n} \sin(\frac{n \pi j}{N+1}) [-a_n + a^{\dagger}_n]
\end{align}
with 
\begin{align}
\omega_n = \omega_0 \frac{2(N+1)}{\pi} \sin(\frac{n \pi}{2(N+1)})
\end{align}
And the Hamiltonian is given by
\begin{align}
H &= \frac{N}{2} \sum_{j} \left[\tilde{\pi}^2_j + \frac{\omega_0^2}{\pi^2} (\tilde{\phi}_{j+1} - \tilde{\phi}_j)^2 \right]
\end{align}
For the case of the hard box, $\hbar \omega_0 = \alpha N$. The form of the redefined fields has the advantage that they can be generalized to the case of a non-uniform potential. We will henceforth work with these redefined fields, dropping the tildes.

\subsection{Non-uniform potential}\label{sec:non-uniform-lat}

How does one construct a lattice boson corresponding to fermions in a non-uniform, confining, potential? The lattice formulation for the hard box case gives us a small hint: we note that the $N$ lattice points in that case correspond to nodes of the Fermi level wavefunction. In particular since the wavefunction is sinusoidal, the lattice points are uniformly dense. For a non-uniform potential, the nodes of the Fermi level wavefunction are denser in regions where the potential is deeper; so if the above hint is true, then the lattice points will also be denser in those regions (we will find this to be indeed the case). In particular, for a non-uniform potential, the lattice must be non-uniform too.

We postulate that the non-uninform lattice is obtained from the uniform one by a coordinate transformation, as follows
\begin{align}
  x=f^{-1}(y), \; y= f(x) \equiv \fr{L}{\pi \hbar N} \int_{x_-}^x p_F(x') dx', \; p_F(x) = \sqrt{2(\vep_F - V(x))}
  \label{non-uniform}
\end{align}
Here  $y= (0,L)$ represent the end-points of the box, which map to the two turning points $x=x_\mp$ given by the zero of the fermi momentum $p_F(x)$\footnote{It is instructive to check: $L= \int_0^L dy = \fr{L}{\pi \hbar N} \int_{x_-}^{x_+} p_F(x') dx' = \fr{L}{\pi \hbar N} \pi \hbar N = L$.} (see Figure \ref{fig:coordinate-transform}).

\begin{figure}[H]
\begin{minipage}{200pt}
\centering
  \includegraphics[scale=.6]{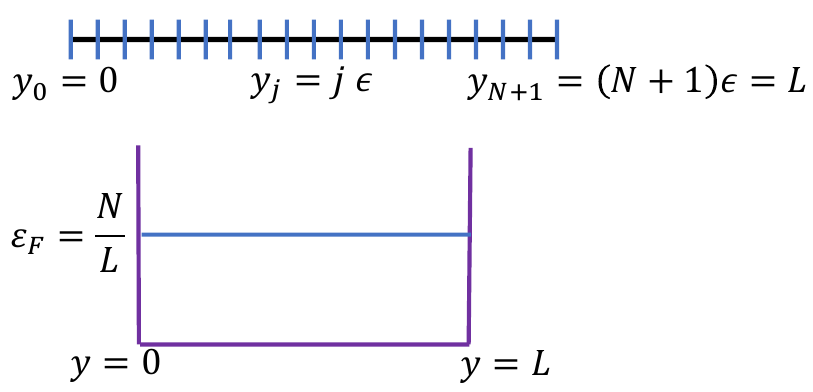}
\end{minipage}
\kern50pt
\begin{minipage}{200pt}
\centering
  \includegraphics[height=3.5cm, width=7cm]{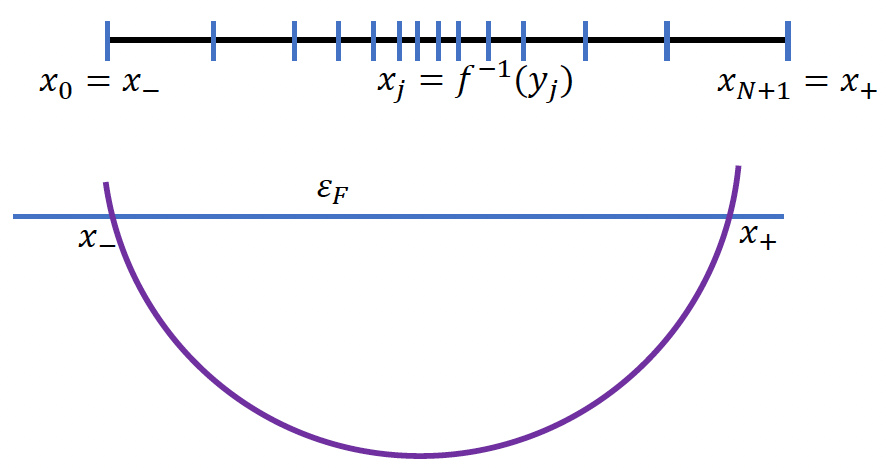}
\end{minipage}
\caption{\footnotesize A schematic description of the non-uniform lattice needed for the coordinate transformation. The left panel shows a uniform lattice similar to the one used in \ref{sec:fixed-lat}, with $y_j = h \ep$, $\ep= L/(N+1)$, along with a cartoon of the fermion potential and the Fermi surface; we call the coordinate $y$. The right panel shows the cartoon of a fermion potential $V(x)$ along with the Fermi surface; the corresponding lattice points $x_j$ are given by a coordinate transformation $y_j= f(x_j)$, where $f(x)$ is defined in \eq{non-uniform}.}
\label{fig:coordinate-transform}
\end{figure}

On the lattice the transformation sends the uniformly spaced points $y_j=  j\ep== j L/(N+1)$, $j=0,1,...,N+1$, with $y_0=0, y_{N+1}=L$, to $x_j =f^{-1}(y_j) =f^{-1}(j \ep)$, $j=0, 1, 2, ...,N+1$, which clearly correspond to a non-uniform set of lattice points. As remarked above, the extreme points map to $x_0= x_-, x_{N+1}= x_+$. 

The lattice boson corresponding to fermions in a non-uniform potential is given by
\begin{align}
  \phi_j &= \sum_{n=1}^{N} \frac{1}{\sqrt{\omega_n}}\sin(\pi n f(x_j)/L)\left[a_n  + a^{\dagger}_n\right] \label{phi-def-nonuniform} \\
	\pi_j &= \sum_{n=1}^{N} \frac{i}{N}\sqrt{\omega_n}\sin(\pi n f(x_j)/L) \left[-a_n + a^{\dagger}_n\right], \label{pi-def-nonuniform}
\end{align}
where
\begin{align}
\omega_n = \omega_0 \frac{2(N+1)}{\pi} \sin(\frac{n \pi}{2(N+1)})
\end{align}
with $\omega_0$ given by the low energy approximation of the exact boson Hamiltonian corresponding to the non-uniform potential (see \eqref{omega0}). This amounts to identifying $\phi(y_j)= \phi(f(x_j))$ and $\pi(y_j)= \pi(f(x_j))$ provided we replace the hard box $\omega_0 = \alpha N / \hbar$ with the general $\omega_0$ from \eqref{omega0}.

The lattice Hamiltonian \eq{lattice-ham-text} (which describes the system at low energies) now reads (by using the fact that $j\ep= y_j = f(x_j)$)
\begin{align}
  & H_{lattice}= \frac{N}{2} \sum_j\left[\pi(f(x_j))^2 + \frac{\omega_0^2}{\pi^2} \left(\phi(f(x_{j+1})) - \phi(f(x_j))\right)^2 \right]
  \label{lattice-ham-nonuniform}
\end{align}
The exact oscillator Hamiltonian with a non-uniform confining potential is given by \eq{bose-ham}; the lattice version of that Hamiltonian can be obtained by (a) inverting the equations \eq{phi-def-nonuniform},\eq{pi-def-nonuniform} to solve for the oscillators $a_n, a^\dagger_n$ in terms of $\phi_j = \phi_{f(x_j)}, \pi_j = \pi_{f(x_j)}$, and (b) substituting these expressions in \eq{bose-ham}. The Hamiltonian \eq{lattice-ham-nonuniform} is the low energy form of that exact lattice Hamiltonian.

The coordinate transformation \eq{non-uniform} has the following invariance property for small intervals $\Delta y= f'(x) \Delta x$
\begin{align}
   x\to y: p_{F,0}(y) \Delta y = p_F(x) \Delta x
  \label{invariance-xy}
\end{align}
where $p_{F,0}(y)= \hbar \pi N/L$ is the Fermi momentum in the $y$-coordinate system (corresponding to the hard box potential).

Note that the set of coordinate transformations with the above invariance property \eq{invariance-xy} form a group. That is, if we make a coordinate transformation $y= f(x)$ and a different one $y= \tilde f(\tilde x)$, both satisfying \eq{invariance}, then $x$ and $\tilde x$ are also related by the same invariance property
\begin{align}
   x\to\tilde x: \tilde p_F(\tilde x) \Delta \tilde x = p_F(x) \Delta x
  \label{invariance}
\end{align}
The above coordinate transformations, in particular \eq{non-uniform}, with the defining property \eq{invariance-xy}, \eq{invariance}, are explained in detail in appendix \ref{app:coord}. In particular, it is shown there that \eq{non-uniform} can be regarded as induced from a canonical (or $W_\infty$) transformation, and it transforms the Fermion potential from a box potential to the desired potential. In fact the more general coordinate transformation \eq{invariance} is also explained in that Appendix in a similar way.

We conclude this subsection by noting that the coordinate transformation \eq{non-uniform} implies that the lattice boson can be regarded as living on a uniform spatial lattice in a metric
\begin{align}
  ds^2= - dt^2 + dy^2 = -dt^2 + f'(x)^2 dx^2 = -dt^2 +\left( \fr{L}{\pi \hbar N}\, p_F(x) \right)^2 dx^2
  \label{metric}
\end{align}

\subsection{Entanglement entropy}\label{sec:bose-EE}

What is the entanglement entropy of an interval $A= [x_1, x_2]$  of the lattice boson?

\subsubsection{Hard box}\label{sec:EE-hard-box}

Let us consider the case of the hard box first. The lattice boson, defined in Section \ref{sec:fixed-lat}, with the Hamiltonian \eq{lattice-ham-text},
can be regarded as the lattice version of a massless relativistic boson. To be precise, as mentioned above \eq{lattice-ham-text}, the exact bosonic Hamiltonian \eq{bose-expand} written in terms of the lattice variables is different from and more complicated that \eq{lattice-ham-text}. However, the ground state of the exact hamiltonian, as remarked above \eq{bose-ground}, is exactly given by the `perturbative' ground state $|0\ran_B$ annihilated by all the $a_k$, $k=1,2,...,N$.\footnote{Note that this is not true of a $\phi^4$ theory, where the Hamiltonian includes a term of the type $(a^\dagger)^4$.} It is easy to see that this is also the ground state of the lattice Hamiltonian \eq{lattice-ham-text}. Thus, the job of finding out the entanglement entropy of an interval $A$ in the ground state of the exact bosonic hamiltonian boils down to the same EE in the ground state of the lattice boson theory with the standard lattice Hamiltonian. 

We use the method described in \cite{Casini:2009sr} to numerically compute the ground state EE of the relativistic lattice boson theory. It involves first evaluating the two-point correlators $X_{jk} = \langle \phi_j \phi_k \rangle$ and $Y_{jk} = \langle \pi_j \pi_k \rangle$\footnote{The notation $\lan ... \ran$ means $\lan 0|...|0\ran$.} where the lattice points are in the region $A$ of interest.
The EE is then given in given in terms of the matrix $C=\sqrt{XY}$, as follows:
\begin{equation}
  S_A = \Tr\left(\left(C+\frac{1}{2}\right)\log(C+\frac{1}{2}) - \left(C-\frac{1}{2}\right)\log(C-\frac{1}{2})\right).
  \label{casini-huerta}
\end{equation}
Our numerical result is shown in Figure \ref{fig:EE-vs-l-hard-box}, with the best-fit curve (up to some unimportant additive constant term) approximately given by
\begin{align}
S_A^{\text{ numerical}} \approx \frac{1}{3} \log(\frac{l}{\ep})
\end{align} 
where $l$ is the length of the interval $A$.
\begin{figure}[H]
\begin{center}
  \includegraphics[scale=0.5]{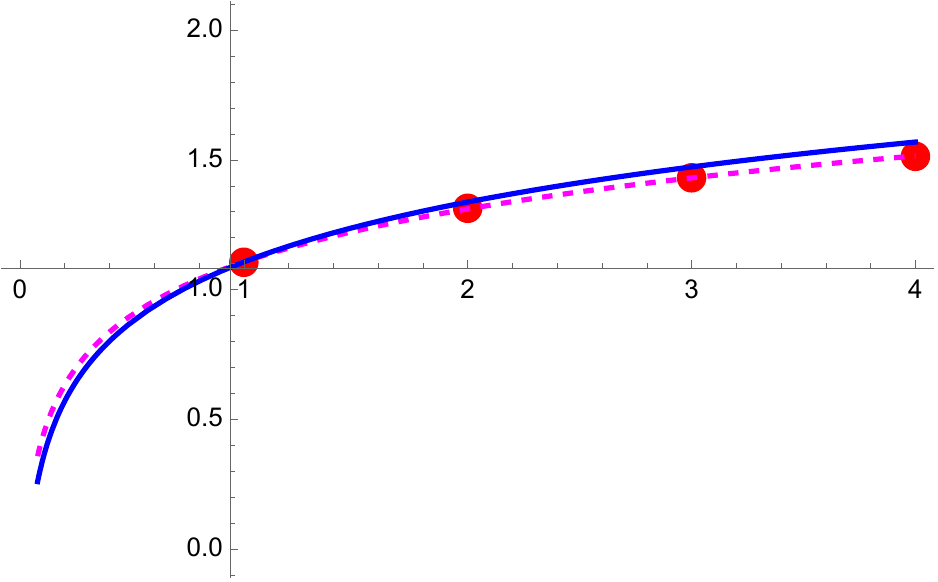}
\end{center}
\caption{\footnotesize Here we have plotted $S_A$ on the vertical axis, and the size $l$ of the interval $A$ (in units of lattice spacing) on the horizontal axis. The values used are $L=50$, $N=100000$, $\ep=L/(N+1)$. The interval is placed at $A=[x_1, x_1 + l]$ where $x_1$ is placed at the midpoint of the hard box. EE is calculated for up to lattice size of $l= 4\ep$. The marked points in red are the numerical values of the EE \eqref{casini-huerta}, the dashed curve (magenta) is the fit to numerical points using the formula $S_A= a + b \log(l/\ep)$ (we get $a=1.11, b=0.30$). The blue curve is the ideal curve $S_A= a + 1/3 \log(l/\ep)$ (we have chosen the same value of $a$).}
\label{fig:EE-vs-l-hard-box}
\end{figure}
\noindent We therefore observe that our computation of EE matches with that of the UV-regulated continuum relativistic massless scalar theory \cite{Holzhey:1994we, Calabrese:2004eu} (which is a two-dimensional CFT with $c=1$) with the short distance cut-off identified with the lattice spacing 
\begin{align}
  S_A= \fr13 \log(\fr{l}{\ep}) + c_1'
  \label{calabrese-cardy}
\end{align}
where $c_1'$ is a non-universal constant that depends on the regularization scheme. Although \eqref{calabrese-cardy} was originally derived \cite{Holzhey:1994we, Calabrese:2004eu} for a boson in a non-compact space, it has been shown in \cite{calabrese:2010he} that it continues to hold when the boson is placed instead in a hard box of length $L \gg l$, and the interval $A=[x_1, x_2]$ is sufficiently far away from the boundaries (e.g. $|x_1|, |x_2| \ll L/2$ when the boundaries of the hard box are at $\mp L/2$).\footnote{Note that the computation of the EE involves placing twist operators at the two points $|x_1, x_2|$. The boundaries lead to image points; thus if the strip $A \times$ the Euclidean time is conformally mapped to an UHP, we are dealing with a four-point function of twist operators. When the interval is sufficiently in the interior, the twist operators have subleading correlators with their images, and hence the result boils down to \eq{calabrese-cardy}. There is a subtle issue if there are relevant operators in the theory of vanishing conformal weight. In $c=1$ boson with a non-compact zero mode there are indeed such operators; however, the lattice boson defined in Section \ref{sec:fixed-lat} excludes the zero mode sector.}

A similar demonstration that the numerical calculation of EE agrees with the continuum-limit result of the boson theory is made in \cite{Callan:1994py} and is remarked in various other papers \cite{Peschel:2002yqj,Eisler:2009vye}. All of this suggests that we can indeed regard the lattice version of a relativistic massless boson as lattice regularization of the corresponding continuum theory with the UV cut-off $\Lambda \propto 1/\ep$. We therefore see that the lattice boson in the hard box problem yields the following ground state entanglement entropy (dropping the unimportant additive constant, and equating $N+1 \approx N$ for large $N$)
\begin{align}
  S_A= \fr13 \log(\fr{(x_2-x_1) N}{L})
  \label{calabrese-cardy-2}
\end{align}
which exactly reproduces the result \eq{SS-v0} for the EE in the ground state of $N$ fermions in a hard box (the coordinate $x$ is to be identified with $\l$).

\subsubsection{Non-uniform confining potential}\label{sec:EE-nonuniform}

As we argued in Section \ref{sec:non-uniform-lat}, in this case the lattice boson is defined as in \eq{phi-def-nonuniform} and \eq{pi-def-nonuniform}. Note that although the lattice representation of the appropriate bosonic Hamiltonian \eq{bose-ham} is complicated and interacting (which boils down to \eq{lattice-ham-nonuniform}, the exact ground state is still given by the state $|0\ran$ annihilated by all the annihilation operators $a_n$ (see remarks above \eq{bose-ground}). Therefore, the correlation matrices $X_{jk} = \langle \phi_j \phi_k \rangle$ and $Y_{jk} = \langle \pi_j \pi_k \rangle$, viewed as a function of the labels $j,k$ of the lattice points remain the same. Noting that these labels are simply related to the $y$-coordinates: $y_j = j \ep$, it is clear that the EE is again given by \eq{calabrese-cardy-2} with the replacement $x \to y$ (since the uniform lattice coordinate is now being called $y$):
\begin{align}
  S_A= \fr13 \log(\fr{(y_2-y_1) N}{L})
  \label{calabrese-cardy-3}
\end{align}
We can now use the coordinate transformation \eq{non-uniform} to cast the EE in terms of the coordinate $x$ of the non-uniform lattice. The interval $A$ is now given by $[x_1, x_2]$ (where $y_1= f(x_1), y_2=f(x_2)$). Assuming that the interval is small enough so that the Fermi momentum $p_F(x)$ does not vary appreciably, we can write
\[
\fr{(y_2-y_1) N}{L}=\fr{N}L f'(x) (x_2-x_1)= p_F(x)(x_2-x_1)
\]
so that \eq{calabrese-cardy-3} reduces to
\begin{align}
  S_A= \fr13 \log(\fr{p_F(x)(x_2-x_1)}{\hbar})
  \label{calabrese-cardy-4}
\end{align}
which agrees with \eq{SS}, up to additive constants which will not be important for us.

\subsection{Exactness and universality of the EE}\label{sec:exact-EE}

It is important to re-emphasize (already mentioned above Equation \eq{bose-ground}) that the state $|0 \ran$ is the universal ground state for {\it all} Hamiltonians \eq{bose-ham}. Thus, the ground state entanglement entropy for all Hamiltonians is the same; namely it is given by \eq{calabrese-cardy-3} in terms of the flat lattice coordinates $y_j$ (in case of the non-uniform potential, the functional form of the EE, written in terms of the non-uniform lattice coordinates $x_j$, changes to \eq{calabrese-cardy-4}). In case of the hard box potential, with Hamiltonian \eq{bose-expand}, the ground state EE is precisely given by \eq{calabrese-cardy-2}, irrespective of whether we consider the Hamiltonian to be only $H_1$ or the full Hamiltonian \eq{bose-expand}.

\subsection{Finite temperature EE}\label{sec:finite-temp}

Now that we have established that the ground state EE agrees between our exact lattice bosonic theory and the fermion theory, one may wonder about such an agreement for excited states. In this subsection we show that in the large $N$ limit, at temperatures which do not scale with $N$, the agreement between the bosonic and fermionic EE continues to hold.

\subsubsection{Bosonic EE}

To see this, note that, as already argued above (see also Section \ref{app:lattice-partn-fn}), at energies or temperatures which do not scale with $N$, the exact lattice boson theory is well described by the Hamiltonian \eq{lattice-ham-text} which is simply the lattice version of a massless relativistic scalar (with $v_F$ playing the role of the speed of light). As mentioned before, we may regard the lattice theory as a c=1 CFT with a lattice cut-off $\ep$; the EE for an interval $A$ of length $l$ at a temperature $1/\beta$ for such a theory\footnote{where in the limit of small $\ep$, $\beta$ is held fixed.} is given by
\begin{align}
  S_A = \fr13 \log\left(\fr\beta{\pi \ep} \sinh(\fr{\pi\ l}\beta )\right)
  \label{finite-temp-EE}
\end{align}

\subsubsection{Fermionic EE}

In Appendix \ref{app:relative-approx} we will show how the ground state EE of the fermion can be understood from an equivalent description in terms of two species of free massless relativistic fermions (arising from particles and holes), but equipped with a built-in UV cut-off given by the Fermi momentum $p_F$. We now argue that at sufficiently low temperatures, even finite temperature EE for non-relativistic fermion theory can be understood in terms of these relativistic fermions. To see this, note that the temperature $T=1/\beta$\footnote{We should distinguish this $\beta$ from the one that appears in the definition \eq{matrix-QM} of the matrix QM.} enters in the EE calculation described in Appendix \ref{app:thomas-fermi} as follows:
\begin{align}
  \lan u(x,p) \ran_\beta &= \sum_n u_n(x,p) f_{FD}(n,\beta)
  \nonumber\\
  u_n(x,p) &= \int d\eta\, \chi_n^*(x+\eta/2) \chi_n(x-\eta/2) \exp[ip\eta/\hbar] \nonumber\\
  f_{FD}(n,\beta) &= \fr1{\exp[\beta (\ep_n - \ep_F)]+1}
  \label{fermi-dirac}
\end{align}
In case of the hard box, $\ep_n = \alpha n^2$ where $\alpha = \hbar^2 \pi^2/(2 m L^2)$. The Fermi-Dirac distribution says that modes $n$ with $(\ep_n - \ep_F) \gg 1/\beta$ do not contribute to the above calculation. It is straightforward to show that for
\begin{align}
  \beta \gg \beta_0 = mL^2/\pi^2
  \label{rel}
\end{align}
the contributing modes are also approximately relativistic. Hence, in this range of temperatures, one can compute even the finite temperature EE for the fermions from CFT techniques, leading to the same formula \eq{finite-temp-EE} as for the bosons.

\section{Remarks about $c=1$}\label{sec:c=1}

As we saw in the previous section, the exact lattice boson formulation discussed in this paper, exactly reproduces the EE of non-relativistic fermions.
We will see below how it, consequently, reproduces the double-scaled expression for the EE \eq{c=1-EE-final}.

In case of the $c=1$ matrix model, (see, e.g. the reviews in \cite{Ginsparg:1993is, Klebanov:1991qa, Polchinski:1994mb}; see also the more recent interpretations of the $c=1$ model in, e.g. \cite{McGreevy:2003kb, Klebanov:2003km, Douglas:2003up})
 the potential $V(\l)$ needs to have a quadratic maximum. Working concretely in the following double well potential \cite{Polchinski:1994mb}
 \begin{align}
   \label{double-well}
V(\lambda) = \frac{1}{4} \lambda^2 (\lambda-2)^2,
\end{align}
the number constraint \eq{num-constr}, after putting $\hbar=1/(\b N)$\footnote{Note that since we are discussing double scaling, we have restored $\b$.}, shows that for large $\beta$, $\epsilon_F$ is small and the Fermi sea sits deep in the well which is centered around $\lambda=2$.\footnote{Choosing the right well is a matter of convention; we might as well have chosen the left well. The second well is irrelevant as long as we ignore tunneling. The tunneling amplitude represents instanton effects which have been discussed recently in \cite{Balthazar:2019ypi, Sen:2019qqg, Balthazar:2019rnh}; we will not discuss these in our paper.} As $\beta$ is reduced, the Fermi sea keeps rising until the Fermi surface hits the local maximum of the potential ($\ep_F \to \epsilon_c = 1/4$\footnote{This is the value of the potential maximum which occurs at $\lambda=1$.}) at some critical value of $\beta = \beta_c= 3/4$ and the theory undergoes a phase transition.

The double scaling limit that maps the matrix model to string theory is defined by
\begin{align}
\label{double-scaling}
N \to \infty, \beta \to \beta_c, \; \mu = - \beta N (\epsilon_F - \epsilon_c)
\end{align}
is held fixed. In this limit, the coordinate $\lambda$ gets rescaled to 
\begin{align}
\tilde{\lambda} = \sqrt{\beta N} (\lambda-1)
\end{align}
so that $\lambda$ approaches $1$ as $N \to \infty$ and we zoom in on the inverted quadratic profile of the potential at the local maximum. The Fermi momentum takes the form
\begin{align} 
P_F(\lambda) &= \sqrt{2 \left(\epsilon_F - \frac{1}{4} \lambda^2 (\lambda-2)^2\right)} \nonumber \\
&= \sqrt{2 \left(\epsilon_F - \frac{1}{4} + \frac{(\lambda-1)^2}{2} - \frac{(\lambda-1)^4}{4} \right)}.
\end{align}
We will now see what the general formula \eq{SS} yields for the $c=1$ matrix model (we again use the value of $\hbar$ in \eq{hbar-n})
\begin{align}
S_A &= \frac{1}{3} \log( 2(\l_2 - \l_1) \beta N \sqrt{2\left(\epsilon_F - \epsilon_c +\frac{(\lambda -1)^2}{2} - \frac{(\lambda-1)^4}{4}\right)}  ) \nonumber \\
&= \frac{1}{3} \log( 2\frac{(\tilde\l_2 - \tilde\l_1)}{\sqrt{\beta N}} \beta N\sqrt{2\left(-\frac{\mu}{\beta N} +\frac{\tilde{\lambda}^2}{2\beta N} - \frac{\tilde{\lambda}^4}{4 \beta^2 N^2}\right)}) \nonumber \\
&= \frac{1}{3} \log(2(\tilde\l_2 - \tilde\l_1) \sqrt{2\left(-\mu +\frac{\tilde{\lambda}^2}{2}  \right)})
\label{c=1-EE-app}
\end{align}
The remarkable thing to note here is that EE is finite even though $N$ is taken to infinity. As shown in \eqref{c=1-EE-final}, this result can be recast in string theory language as 
\begin{align}
S_A &= \frac{1}{3}\left( \log\left( \frac{x_2 - x_1}{l_s g_s}\sinh^2(x_0/l_s)\right)\right)
\end{align}
The above derivation, which goes through in the fermion formulation, is equally valid in the lattice boson formulation. This is because the lattice boson reproduces the first step of the above equation, which then goes through all the remaining steps. It also turns out that although $N$ is taken to infinity, the double-scaled exact boson theory still retains a lattice structure with finite lattice spacing given by $g_s l_s$. We discuss this a bit more in Section \ref{subsec:c=1-finiteness} but leave its in-depth details to our upcoming work in \cite{GM-progress2}.

\paragraph{D0 branes} As we detail in the Concluding section \ref{sec:conclusion}, there is a likely interpretation of our exact bosonic field in terms of creation and annihilation of D0 branes in the more recent type 0B interpretation of the $c=1$ model \cite{McGreevy:2003kb, Klebanov:2003km, Douglas:2003up}). 

\section{Proof of finiteness of EE}\label{sec:finite}

We first discuss the case of finite $N$ for both fermions and our equivalent bosons. After that, we tackle the double-scaled $c=1$ theory separately.
\subsection{Fermions}
The EE of a region $A$ in a non-interacting fermionic theory is given primarily by the second cumulant of the particle number distribution
$V_2$ as shown in \eqref{sa-v2}. When computing the Fermi ground state EE, the formula \eqref{v2-result} becomes
\begin{align}
 S_A &= \frac{\pi^2}{3} \left( \int_A dx \sum_{n=0}^{N-1} |\chi_n(x) |^2-  \int_A dx \int_A dx' | \sum_{n=0}^{N-1} \chi_n^*(x) \chi_n(x') |^2\right).
\end{align}
Since the single particle energy eigenfunctions $\chi_n(x)$ do not have any singularity and the sum over modes $n$ is finite, EE is bound to be finite. 

\subsection{Bosons}
For bosons, computation of EE involves first evaluating the two-point correlators $X_{jk} = \lan \phi(x_j) \phi(x_k) \ran$ and $P_{jk} = \lan \pi(x_j) \pi(x_k) \ran $ for all points inside the interval of interest $A$. Then, one constructs the matrix $C = \sqrt{XP}$ and plugs it into the following expression to obtain EE.  
\begin{align} 
S_A = \tr ((C+1/2)\log(C+1/2) - (C-1/2)\log(C-1/2))
\end{align}
In our equivalent bosonic theory, space is discretized since the total number of modes is $N$ and this implies that there are a total of $N$ lattice points. Each of the ground state correlators 
\begin{align}
X_{jk} &= \sum_{k=1}^N \frac{1}{2 \omega_k} \psi_k(x_j) \psi_k^*(x_j) \nonumber \\
P_{jk} &= \sum_{k=1}^N \frac{\omega_k}{2} \psi_k(x_j) \psi_k^*(x_j) 
\end{align}
are finite since the mode sum is finite. The dimension of matrix $C$ is also finite since the number of lattice points inside the region $A$ has to be less than the total number of lattice points $N$. Furthermore, the eigenvalues of the matrix $XP$ are constrained to be greater than $1/4$ \cite{Casini:2009sr}. Since there is no source of divergence anywhere in this calculation, we see that EE is finite for bosons as well.

\subsection{Double-scaled $c=1$} \label{subsec:c=1-finiteness} 
In the previous two subsections, the finiteness of EE was explained as a result of finiteness of $N$. In the fermionic theory, finiteness of $N$ resulted in a finite sum over the occupied levels and in our exact boson theory, it resulted in a finite lattice spacing of the order $1/N$. So, one would naively think that in the strict $N$ taken to infinity limit of the double-scaled $c=1$ model, EE from both the fermionic theory and the exact boson theory might be divergent. However, that turns out to not be the case.

In the fermionic theory, one does encounter divergent integrals while computing EE using the cumulant method but the divergences are shown to precisely cancel out in \cite{Hartnoll:2015fca}, yielding a finite result. In our exact boson theory, we find to our surprise that even though $N$ is taken to infinity, our exact boson still retains a (non-uniform) lattice structure with an effective finite lattice spacing given by $g_s l_s$.  

To see this lattice structure, one has to perform a coordinate transformation from the uniform hard box coordinate $\bar{\lambda}$ to the double-scaled $\tilde{\lambda}$ coordinate as described in Section \ref{sec:non-uniform-lat}.
\begin{align}
\frac{N \pi}{L} d\bar{\lambda} &= \frac{p_F(\tilde{\lambda})}{\hbar} d\tilde{\lambda} \nonumber \\
&= \sqrt{{\tilde{\lambda}}^2 - 2 \mu} \;d \tilde{\lambda} \nonumber \\
&= 2 \mu \sinh^2x \; dx
\end{align}
This suggests that in the $x$ coordinate, the lattice spacing is given by
 \begin{align}
 \delta x_j = x_{j+1} - x_j &= \frac{N \pi }{2 \mu \sinh^2x_j L} (\bar{\lambda}_{j+1} - \bar{\lambda}_j) \nonumber \\
 &= \frac{\pi g_s l_s}{\sinh^2x_j} 
 \end{align}
 where we have used the fact that $2 \mu = 1/ g_s l_s$. 
 \begin{figure}[H]
\begin{center}
  \includegraphics[scale=0.5]{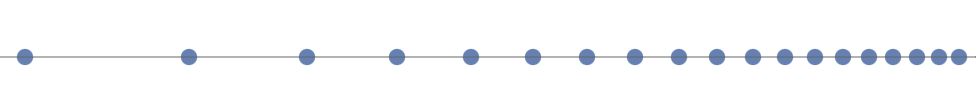}
\end{center}
\caption{\footnotesize This is a sketch of how the lattice structure of our double-scaled $c=1$ exact boson theory looks like for $x >>1$ where the lattice is well-approximated by $x_j \sim g_s l_s \log(j)$}
\label{fig:EE-vs-l-hard-box}
\end{figure}
 The lattice structure is non-uniform, with the effective lattice spacing given by $g_s l_s$. It also turns out that the bosonic mode integral in the normal mode expansion of our lattice boson field (where the sum over $k$ is replaced by an integral over $\tilde{k} \equiv k/N$ with $N$ taken to infinity) is finite since the integration range is finite ($\tilde{k} \in (0,1)$) and the integrand is non-singular. More details on these and our double-scaled $c=1$ exact boson theory will appear in \cite{GM-progress2}.

\section{Issues with the density variable description}\label{sec:prob}

As mentioned before, a natural continuum description of a large dimensional matrix would appear to be its eigenvalue density
\begin{align}
  \label{rho-lam}
  \rho(\l)= \sum_{i=1}^N \delta(\l - \l_i)
\end{align}

\subsection{Trace identities}

Indeed, in the context of the zero dimensional random matrix model, in the classic paper \cite{Brezin:1977sv} the (path) integral for the singlet ($U(N)$ invariant) sector of a large dimensional matrix model is expressed in terms of a path integral over $\rho(\l)$. In the large $N$ limit, the latter path integral is dominated by a saddle point value of $\rho(\l)$ (the `Wigner semicircle distribution' in case of the Gaussian matrix model).\footnote{The precise saddle point solution is $\rho(\l)= \frac1{2\pi}\sqrt{4-\l^2} \theta(4 - \l^2)$ \cite{Brezin:1977sv}.} Expectation values of $U(N)$ invariant quantities, such as the traces
\begin{align}
  \label{moments}
  a(p)\equiv \tr M^p = \sum_{i=1}^N \l_i^p = \int d\l\, \rho(\l) \l^p
\end{align}
are correctly reproduced in the strict $N\to \infty$ limit by this saddle point value.

The main conceptual issue with the density description has to do with the fact that a function $\rho(\l)$ (subject to only being positive semi-definite) has infinitely more information than the matrix itself, for any finite $N$, however large. These are encapsulated by the {\it\ub{trace identities}} (also known as Cayley-Hamilton identities, see, e.g. \cite{atiyah1994introduction}) which relate moments of order higher than $N$, tr $M^{N+p}, p=1,2,...$ to lower moments tr $M^{N-q}, q=0,1,...,N-1$. For a simple example, let us consider $N=2$, i.e. a 2 $\times$ 2 matrix $M$, with two eigenvalues $\l_1, \l_2$. It is clear in this case that the moments tr $M$ and tr $M^2$ are enough to determine $\l_1, \l_2$; hence tr $M^3$ and all higher moments must be expressible in terms of tr $M$ and tr $M^2$; e.g. the first trace identity is
\begin{align}
  \tr M^3 = \frac32 \tr M\ \tr M^2 - \frac12 (\tr M)^3
  \label{cayley-n=2}
\end{align}
If we express this equation in terms of the eigenvalue density through \eq{moments}, it gives a rather nontrivial identity which must be satisfied by $\rho(\l)$ 
\begin{align}
\int d\l\, \rho(\l) \l^3 = \frac32 \int d\l\, \rho(\l) \l^2 \int d\l'\, \rho(\l') \l'  - \frac12 \int d\l\, (\rho(\l) \l )^3
  \label{cayley-n=2-rho}
\end{align}
There are infinitely more such identities involving $\tr M^{2+n}, n> 0$. A general function $\rho(\l)$ (even with the condition $\rho(\l) \ge 0$) does not satisfy these constraints as its moments are {\it a priori} all independent. Unless one incorporates all these trace constraints into a path integral over $\rho(\l)$, such a path integral cannot be a correct translation of the random matrix integral.

A different, slightly more mathematical, way of saying the same thing is the following. Identities like \eq{cayley-n=2-rho}, even for large $N$, cannot be satisfied by a smooth function $\rho(\l)$; indeed they should not be, since the quantity \eq{rho-lam} is not a {\it function}, it is a distribution. Thus, an appropriate translation of the random matrix integral should be in terms of an integral over distributions.

One might argue that in the strict $N\to \infty$, the trace identities should become vacuous and hence there should not be any problem in a description in terms of smooth functions $\rho(\l)$. However, it is not {\it a priori} clear whether this is a reasonable viewpoint when one wishes to explore subleading corrections in $1/N$. \footnote{It {\it would be} reasonable if the description in terms of smooth functions $\rho(\l)$ leads to a violation of the trace identities which is weaker than powers of $1/N$; this question is currently being investigated in \cite{GM-progress1}. See also the remarks in the next paragraph on the ``Stringy exclusion principle''.}

The above discussion was for zero dimensional matrices. The same observations apply also to matrix quantum mechanics, where the density variable description is encapsulated in the collective variable theory of \cite{Das:1990kaa}. Once again, the transformation from the matrix variable to the eigenvalue density or its moments has the problem that the latter do not satisfy the trace identities and hence has infinitely more degrees of freedom.

We note here that the effect of finite $N$ shows up in a modification of Poisson bracket to Moyal bracket and ordinary product to star product in a description of matrix QM in terms of a second quantized Wigner phase space distribution \cite{Dhar:1992hr, Dhar:1992rs, Das:1991uta} (this is closely related to the appearance of $W_\infty$ algebra in this system).

It is perhaps not inappropriate to mention a similar problem for higher dimensional matrix theories too. In the context of AdS/CFT, the consequence of a finite rank $N$ of super Yang Mills has sometimes been observed to result in nonperturbative bounds on appropriate charges, a result typically called the ``Stringy exclusion principle'' (see, e.g., \cite{Maldacena:1998bw} in the context of AdS$_3$/CFT$_2$, and \cite{McGreevy:2000cw, Balasubramanian:2001nh} in the context of giant gravitons). We have seen above that the hint of an appropriate alternative to the eigenvalue density description is in fact provided by the story of giant gravitons.

\subsection{Moments} \label{sec:moments}
In MQM (or its equivalent fermionic theory), one can perform a following exact calculation of moments 
\begin{align}\label{ip-MQM}
I_p =\langle \frac{1}{N} \tr M^p \rangle = \frac{1}{N} \sum_{n=0}^{N-1} \int d\lambda \;  \lambda^p | \chi_n (\lambda) |^2 .
\end{align}
This will have a definite answer in terms of $N$. The $4$th moment in the SHO potential, for e.g., 
\begin{align} 
I_4 = \frac{\hbar^2}{m^2 \omega^2} \left( \frac{N^2}{2} + \frac{1}{4} \right)
\end{align}
with the identification $\hbar = 1/(\beta N)$ is given by
\begin{align} \label{i4-SHO}
I_4 = \frac{1}{\beta^2 m^2 \omega^2} \left( \frac{1}{2} + \frac{1}{4 N^2} \right)
\end{align}
Let us now investigate whether the collective field theory corresponding to the SHO potential MQM can reproduce the above result. The action (written conveniently in terms of $\varphi(x,t)$ where $\rho_c(x,t) = \beta N \partial_x \varphi(x,t)$) is given by 
\begin{align} \label{coll-action}
S_c = N^2 \int dt \; dx \; \left\{ \frac{1}{2} \dot{\varphi}^2 \partial_x \varphi + \frac{\pi^2}{6} (\partial_x \varphi)^3 + V(x) \partial_x \varphi \right\}, 
\end{align}
with $V(x)=m \omega^2 \lambda^2/2$. The equivalent collective variable calculation to be done is 
\begin{align}
I_p = \frac{1}{Z} \int \mathcal{D} \varphi \left( \frac{1}{N} \int dx \; x^4 \rho_c(x,t)  \right) e^{i S_c}.
\end{align}
Since $S_c \sim N^2$, the path integral in the large $N$ limit is dominated by the following saddle point solution to the equation of motion of $S_c$.
\begin{align}
\rho_0 (x) = \beta N \partial_x \varphi_0(x) = \frac{\beta N}{\pi} \sqrt{2m \left(\frac{\omega}{\beta} - \frac{m \omega^2 x^2}{2}\right)}
\end{align}
We can verify that this is the correct saddle point solution by using it to calculate the leading contribution to $I_4$
\begin{align}
I_4^{leading} &= \frac{1}{N} \int dx \; \rho_0(x,t) x^4 \nonumber \\
&= \frac{m \omega \beta}{\pi} \int dx \; x^4 \sqrt{\frac{2}{\beta m \omega} - x^2} = \frac{1}{2 \beta^2 m^2 \omega^2},
\end{align} 
which matches with what is expected from \eqref{i4-SHO}. The subleading contribution from the collective theory is divergent since it involves the calculation of loop integrals such as the following tadpole diagram, which is quadratically divergent.
\begin{figure}[H]
\begin{center}
  \includegraphics[scale=0.35]{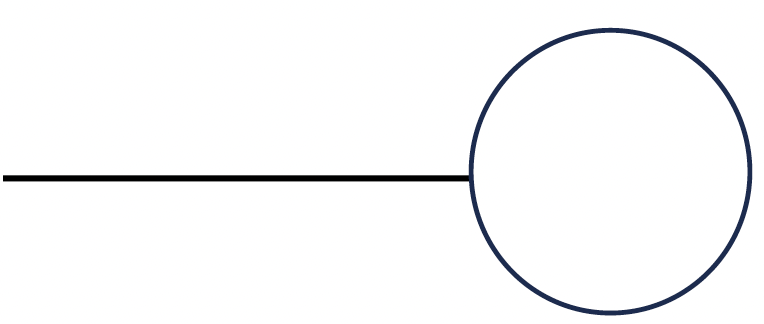}
\end{center}
\caption{\footnotesize The tadpole vertex represents the cubic interaction term $(\partial_x \varphi)^3$ in the collective action \eqref{coll-action}. This diagram has a $\Lambda_{UV}^2$ quadratic divergence, where $\Lambda_{UV}$ is the UV momentum cut-off.}
\label{fig:tadpole}
\end{figure}
 
\subsection{Entanglement entropy}

As explained in the introduction in Section \ref{sec:intro}), the formula for the entanglement entropy \eq{SS} or \eq{SS-v0} indicate the presence of a short distance cut-off related to the finite depth of the Fermi sea: a fact not explicable from the density variable description. In the double scaled $c=1$ theory, the short distance cut-off implied by the EE is \eq{ls-gs}, which is not natural from the density variable or the dual tachyon field of string theory.

Since small fluctuations of the density variable near the Fermi surface behave like a massless relativistic boson, the result for the EE of a small interval is expected to have the usual logarithmic divergence, as in \eq{relativistic}. It is important to mention here the recent paper \cite{Das:2022nxo}, which has a nice description of the finite result for the {\it fermionic} EE \footnote{analogous to our equation \eq{sa-v2} (see their equation (1.5))}, in terms of collective variables, in case the matrix potential is zero. This paper does not compute the {\it bosonic} EE which would be obtained, e.g., by applying \eq{casini-huerta} to the collective variable theory. By contrast, in the exact lattice bosonization described in this paper, a finite result for the EE is obtained directly from the bosonic theory, applying \eq{casini-huerta}.

\subsection{Resolution of the above problems with the Exact lattice boson}\label{sec:resolution}

\begin{enumerate}

\item The trace identities are automatically satisfied in the fermionic path integral with a fixed, finite number $N$ of fermions. Since the change of variables between the fermionic and bosonic variables is exact (in particular, the fermionic path integral can be transformed to a path intgegral in terms of the exact boson variables), the trace identities continue to hold in the bosonic formulation. The main point here is that the lattice boson $\phi_j$ has only $N$ degrees of freedom, just like the finite number of matrix eigenvalues (in contrast with the collective variable $\rho(\l)$ which is a function of a continuous variable). 

\item We observed in Section \ref{sec:moments} that the collective theory calculation of MQM moments $\eqref{ip-MQM}$ is able to reproduce only the leading contribution but not the subleading contribution (which turns out to be divergent). However, the equivalent calculation of moments in our exact boson theory (using the exact state map and the exact operator correspondence) is given by
\begin{align}
I_p &= \frac{1}{N} \bra{F_0} \int d\lambda \; \lambda^p \psi^{\dagger} \psi(\lambda) \ket{F_0} \nonumber \\
&= \frac{1}{N} \sum_{n=0}^{\infty} \bra{F_0} \psi^{\dagger}_n \psi_n \ket{F_0} \int d\lambda \; \lambda^p |\chi_n(\lambda)|^2 \nonumber \\
&= \frac{1}{N} \sum_{n=0}^{\infty} \bra{0} \sum_{k=1}^N \delta\left(\sum_{i=k}^N a_i^{\dagger} a_i -n +N - k\right) \ket{0} \int d\lambda \; \lambda^p |\chi_n(\lambda)|^2 \nonumber \\
&= \frac{1}{N} \sum_{k=1}^N \int d\lambda \; \lambda^p |\chi_{N-k}(\lambda)|^2.
\end{align}
This is inherently finite since our bosonic mode sum $k$ comes with a built-in cut-off $N$ and it matches exactly with the MQM result \eqref{ip-MQM}. In particular, the $4$th moment calculation of the SHO potential in our exact boson theory gives both the leading and subleading contribution correctly
\begin{align}
I_4 = \frac{1}{\beta^2 m^2 \omega^2} \left( \frac{1}{2} + \frac{1}{4N^2} \right),
\end{align}
which matches with the MQM result \eqref{i4-SHO}.
  
\item As shown in detail in the previous sections, in particular Section \ref{sec:finite}, the finiteness of the EE for finite $N$, however large,  is related in the fermion theory  to a finite Fermi momentum. This has a natural understanding in terms of the exact lattice boson description. Due to the fact that the number of bosonic oscillators is finite ($=N$), the number of lattice points in which the lattice boson lives is also $N$, leading to a built-in short distance cut-off $\sim 1/N$. The finiteness in the double-scaled theory from the exact boson viewpoint comes by noting that the lattice boson theory reproduces the fermionic/MQM formula for the $c=1$ EE prior to taking the double scaling limit; hence the demonstration of finiteness of the double scaled limit of the EE works equally well for the fermions and for the bosons, as mentioned previously. The finite lattice spacing of the double-scaled $c=1$ exact boson also supports this fact, as explained in Section \ref{subsec:c=1-finiteness}.

\end{enumerate}

\section{Concluding remarks and outlook}\label{sec:conclusion}

In this paper, we have discussed an exact bosonization of non-relativistic fermions which overcomes a number of limitations of the collective field formulation of matrix QM, in particular it substitutes a natural cut-off in the fermionic EE provided by the Fermi surface, by a lattice cut-off in the bosonic theory. We showed that the latticized exact boson also explains the double-scaled EE of the $c=1$ theory.

It is important to find the significance of this new bosonic formulation in case of the $c=1$ model.\footnote{We thank Ashoke Sen and Juan Maldacena for crucial discussions on this topic.} In the more recent interpretation of the $c=1$ model in, e.g. \cite{McGreevy:2003kb, Klebanov:2003km, Douglas:2003up}, the particle and hole states have been interpreted as D0 brane states (although the interpretation of the {\it hole} states as D0 branes is somewhat indirect). As we saw earlier in the text, the $a_k, \ad_k$ oscillators, or equivalently, the lattice boson fields $\phi(x_j)$, $\pi(x_j)$, describe particle-hole pairs. A particle-hole pair in which the particle sits on the top of the Fermi surface and the hole is $k$ levels down, can be regarding essentially as a hole state. Similarly a particle-hole pair in which the hole is created on the Fermi surface, can be regarded essentially as a particle state. Thus, the $a_k, \ad_k$ oscillators, and the lattice boson fields $\phi(x_j)$, $\pi(x_j)$, have a natural interpretation in string theory in terms of creation and annihilation of D0 branes, rather than perturbative string modes like the tachyon.\footnote{Note that in the somewhat related context of giant gravitons (see Appendix \ref{app:schur} and \cite{Dhar:2005su, Dhar:2005fg}), the $a_k, \ad_k$ oscillators create giant gravitons rather than gravitons which are the perturbative quanta.} This perhaps makes the appearance of the string coupling in the EE formula \eq{c=1-EE-final} more natural, since the D0 brane has mass $\sim \fr{1}{g_s l_s}$, which, furthermore, has an interpretation in terms of the Fermi energy (see Section 4 of \cite{Klebanov:2003km}, where the mass of the D0 brane, on top of the potential, is interpreted as the energy $N\mu$ required to lift it there from the Fermi surface). Note that the tachyon field, related to the eigenvalue density, at the quadratic level, does not know about the string coupling, which is related to the fact that small ripples on the Fermi surface do not know about the top of the potential. We hope to develop these arguments more in \cite{GM-progress2}. 

It is tempting to speculate whether the appearance of the string coupling in the EE formula \eq{c=1-EE-final} is related in any way to the Bekenstein-Hawking entropy which involves the Planck scale \cite{Susskind:1994sm}. We hope to explore this further. Another important point to note is that the usual dual formulation of $c=1$ in terms of the string theory tachyon field works very well for asymptotic observables like the S-matrix \cite{Sengupta:1990bt, Mandal:1991ua, Moore:1991sf, Polchinski:1991uq}, including nonperturbative contributions \cite{Balthazar:2019rnh, Sen:2019qqg}; on the other hand, our work seems to suggest that `local observables' like the entanglement entropy seem to require a different dual bosonic description.\footnote{This point has been made to us by A. Sen.}

As mentioned before, our exact lattice bosonization for large $N$ matrix quantum mechanics can be regarded as a toy model of an emergent geometry with an intrinsic, well-defined, granularity. Somewhat similar issues as discussed in this paper arise also in the case of Random matrix integrals (the ``$c=0$" model); we hope to report on this soon \cite{GM-progress1}.

\subsection*{Acknowledgments}

It is a great pleasure to thank Avinash Dhar, Abhijit Gadde, Shiraz Minwalla, Onkar Parrikar, Suvrat Raju, Spenta Wadia, and especially Sumit Das, Juan Maldacena and Ashoke Sen for crucial discussions. We also thank Sumit Das for his comments on a draft of this paper. G.M. would like to thank organizers of the meetings entitled ``Large-N Matrix Models and Emergent Geometry'' (September 4-8, 2023, ESI, Vienna, Austria) and ``Aspects of CFTs'' (January 8-11, 2024, IIT Kanpur, India) for the opportunity to present a preliminary version of the work, and would like to thank participants of both meetings for many stimulating discussions. We are also grateful to the anonymous referee for the insightful comments and questions which led to enhanced clarity of our presentation in the final version of the paper. We acknowledge support from the Quantum Space-Time Endowment of the Infosys Science Foundation. 

\appendix

\section{Large $N$ matrix quantum mechanics, fermi fluid, and EE}\label{app:matrix-QM}

In the following we will use the convention and notations of \cite{Polchinski:1994mb}.\footnote{$\beta$ below is analogous to the prefactor $1/g_{\rm YM}^2$ in Yang-Mills action.} 
\begin{align}
  S= \beta N \int dt \{\frac12  \Tr (\dot M)^2 - \Tr V(M) \}
  \label{matrix-QM-a}
\end{align}
After a similarity transformation, the dynamics of the eigenvalues of $M$ is described by $N$ free non-relativistic fermions trapped in a $V(\lambda)$ potential.
\begin{align}
  H \to \beta N H'', \;
  H''=\sum_{i=1}^N h(\frac\del{\del\l_i}, \l_i), \kern20pt h(\frac\del{\del\l}, \l)=  -\frac1{(\beta N)^2} \frac{\del^2}{\del \l^2} + V(\l) 
\end{align}
Note that $(\beta N)^{-1}$ plays the role of $\hbar$ in $H''$:
\begin{align}
  \hbar = \frac1{\beta N}
  \label{hbar-n}
\end{align}

\subsection{Fermi fluid droplet}

In the large $N$ limit \eq{large-N} $\beta$=fixed, $N\to\infty$, we have $\hbar\to0$.\footnote{In the following, unless otherwise stated, we will fix $\b=1$ (we will bring it back when we discuss double scaling).} In this limit, states of the fermion theory $|F\ran$ can be represented by a smooth phase space distribution $u(\l,p)$ obtained by taking expectation value of the second quantized Wigner distribution (using notations from \eq{second-q}) \cite{Dhar:1992hr,Dhar:1992rs,Das:1991uta}
\begin{align}
\hat U(\l, p)=\int d\eta\ \Psi^\dagger(\l+ \eta/2) \Psi(\l- \eta/2) e^{i p \eta/\hbar},\quad  u(\l, p)=  \lan F|\hat U(\l, p) |F \ran 
\label{wigner-exp}
\end{align}  
In the large $N$ limit, $u=1$ inside a region $R$ occupied by fermions in the state $|F\ran$, and $u=0$ outside $R$. In other words, $u$ represents a fermi fluid ``droplet'' in phase space, of unit height as demanded by the Pauli exclusion principle. In the ground state $|F_0 \ran$, the fermions fill up the lowest energy single-particle eigenstates up to the fermi surface $h(\l,p)=\epsilon_F$; hence the droplet extends up to the fermi surface:
\begin{align}
  u(\l,p)= \theta(\ep_F - h(\l,p))=\Theta\left(\epsilon_F - \frac{p^2}{2} - V(\lambda)\right)
  \label{droplet}
\end{align}
The expectation value of Weyl-ordered operators is obtained as an integral  
\begin{align}
\lan F| \int d\l \Psi^\dagger(\l) O_W(\l, \del_\l) \Psi(\l)|F \ran=
\int \fr{d\l dp}{2\pi h} u(\l,p) O(\l,p)
\label{exp-val}
\end{align}
Taking the example of the identity operator in \eq{exp-val}, we get the fermion number constraint
\begin{align}
  N = \int \frac{dp}{2 \pi \hbar} \int d\lambda\, u(\l,p)= \fr1{\pi\hbar} \int d\lambda\, P_F(\l)
  \label{num-constr}
\end{align}
Here, the Fermi momentum $P_F(\l)$ is obtained by solving for $p$ the equation for the Fermi surface $\ep_F= h(\l,p)= p^2/2 + V(\l)$, giving
\begin{align}
  p=\pm P_F(\l), \quad P_F(\l) \equiv \sqrt{2(\ep_F - V(\l))}
  \label{positive-pf}
\end{align}
Intuitively, equations like \eq{exp-val} and \eq{num-constr} mean that each fermion (at $(\l,p)$) occupies a phase space volume of $\frac{\Delta p \Delta \l}{2\pi \hbar}$.

\subsection{EE from the fluid droplet picture}\label{app:thomas-fermi}

The material of this section is based on \cite{Smith:2020gfl, Das:2022mtb}. The notation $x$ in this subsection should be identified with $\l$ of the previous subsection.

The EE of an interval $A \in {\bf R}$ for $N$ one dimensional non-relativistic free fermions is given, in the large $N$ limit, in terms of the second cumulant:
\begin{align}
  S_A =\frac{\pi^2}{3} V_2,\; V_2 \equiv  \left( \lan N_A^2 \ran - \lan N_A \ran^2 \right)
  \label{sa-v2}
\end{align}
where the local number operator $N_A$ is given by
\begin{align}
  N_A \equiv \int_A dx\; \Psi^\dagger (x) \Psi(x)
  \label{n-a}
\end{align}
We will be concerned with expectation values in the Fermi ground state $|F_0 \ran$, i.e. the filled Fermi sea: $ \lan ... \ran \equiv \lan F_0 | ... | F_0 \ran $. It can be shown that (in the following, expectation values are understood to be taken at the Fermi ground state)
\begin{align}
  V_2=  \lan N_A \ran -  \int_A dx \int_A dx' \lan \Psi^\dagger (x) \Psi(x') \ran^2
  \label{v2-result}
\end{align}
Let us translate this to the droplet language. By inverting the relation \eq{wigner-exp} between the fermion bilinear and the phase space density, and using \eq{droplet} we get
\begin{align}
  &\lan \Psi^\dagger(x) \Psi(x') \ran = \int \frac{dp}{2\pi\hbar}\ u(\frac{x+x'}{2}, p)\ e^{-i p (x-x')/\hbar} = \frac1{\pi(x'-x)}\sin\left(\frac{x'-x}{\hbar}\ P_{_F}(\frac{x+x'}{2}) \right)
  \label{inverse-droplet}
\end{align}
The expression \eq{SS} for the entanglement entropy follows by substituting the above into \eq{v2-result} and \eq{sa-v2} (see \cite{Smith:2020gfl, Das:2022mtb}).

In the simple example of a hard box (an infinite square well) potential (represented by \eq{hard-box}), the droplet looks like in Figure \ref{fig:potential}(c). The Fermi energy is $\ep_F =\fr{\hbar^2 (N \pi/L)^2}{2}$, hence the Fermi surface is given by
\begin{align} \label{beta-fermiE-0}
p=\pm  P_F, \quad P_F = \sqrt{2\epsilon_F} = \hbar N\pi/L.
\end{align}
In this case, the formula \eq{SS} becomes
\begin{equation} \label{SS-hardbox}
S_A = \frac{1}{3}\, \log\left(\frac{2 P_F\, (\lambda_2 - \lambda_1)}{\hbar}\right)= \frac{1}{3}\, \log\left(\frac{N}{L}(\lambda_2 - \lambda_1)\right).
\end{equation}

\subsubsection{Note on the sign of the momentum $p$}\label{app:p>0}

We make a remark here about the range of the momentum $p$ for the fluid droplets. Consider fermions in a hard box, with wavefunctions proportional to $\sin(n\pi x/L)$, where $n$ is a positive integer (see, e.g., \eq{hard-box}); the Fermi sea consists of $n=1,2,..,N$. On the face of it, this seems to be in conflict with the fact in the above discussions (see, e.g. \eq{beta-fermiE-0}), the Fermi sea comprises all momenta from $p=- P_F$ to $p= P_F$, i.e. momenta which satisfy $p^2/(2m) \le \ep_F$. 

The apparent conflict above arises from a purported identification of $p$ with $n\pi/L$. The resolution of the problem is that this identification is incorrect since $\sin(n\pi x/L)$ is {\it not} an eigenfunction of the momentum operator $-i \hbar \del/\del x$. More simply stated, in a hard box, we have only standing waves and no travelling waves. What is the meaning, then, of the momentum $p$? It is that every positive integer value of $n$, or the corresponding energy eigenvalue $\ep_n$, corresponds to an orbit $h(x,p)= \ep_n$ in the semiclassical phase space. In the preceding discussions in this section, the momentum refers to the $p$-coordinate of a point of the orbit for a given value of $x$. Since $h$ is quadratic in $p$, each $n$-th orbit is reflection-symmetric about $p=0$.

In fact, more quantitatively, the Wigner distribution for the $n$-th eigenfunction $\psi_n(x)$
\begin{align}
  u_n(x,p) = \int d\eta \, \psi_n^*(x+\eta/2) \psi_n(x-\eta/2) e^{i p \eta/\hbar}
 =  \int d\eta \, \psi_n(x+\eta/2) \psi_n(x-\eta/2) e^{i p \eta/\hbar}
  \label{single-wigner}
\end{align}
satisfies $u(x,p)= u(x,-p)$. To show this, note that the integral representation for $u(x,-p)$ becomes that of $u(x,p)$ under a change of the integration variable $\eta \to -\eta$; we have assumed here that the wavefunction $\psi_n(x)$ is real, which is true in case of the hard box or in any confining potential. 

There is yet another way of understanding the emergence of negative momenta for the hard box. It is that linear combination of standing waves can be approximated by travelling waves away from the boundaries. More precisely, we can map a pair of standing waves $\sin(m\pi x/L)$ with even and odd mode numbers $m=2n$ and $m=2n-1$ to a pair of left- and right-moving wave on a circle; the trick is to use the identity (for $n=1,2,...,N'=N/2$)
\begin{align}
  & A_{2n} \sin(2n\pi x/L) + A_{2n-1}  \sin((2n-1)\pi x/L)
  \nonumber\\
  &= A_{2n} \sin(2n\pi x/L) + A_{2n-1}  \left( \sin(2n\pi x/L) \cos(\pi x/L) -
  \cos(2n\pi x/L) \sin(\pi x/L) \right) \nonumber\\
  & \approx A_{2n} \sin(2n\pi x/L) - A_{2n-1}  
  \cos(2n\pi x/L)\nonumber\\
  & =-\fr{i}{2} \Big((A_{2n} - i A_{2n-1}) \exp[i 2n\pi x/L]
  - (A_{2n} + i A_{2n-1}) \exp[-i 2n\pi x/L]\Big)
  \label{p-neg}
\end{align}
In the second step we have assumed that $x$ is sufficiently far from the boundaries of the hard box: $|x-L/2| \ll L/2$, so that $\sin(\pi x/L)\approx 1$, $\cos(\pi x/L)\ll 1$.  Therefore, sufficiently far from the boundaries of a hard box, each successive pair of standing waves of wavenumbers $(2m \pi/L, (2m-1)\pi/L)$ can be regarded as approximately equivalent to a pair of left- and right-moving waves of wavenumbers $(2m \pi/L, -2m\pi/L)$, $m=\pm 1, \pm 2, ..., \pm N'$. In particular if we choose $A_{2n}= -i A_{2n-1}$ we pick up a purely positive momentum (the result $u(x,p)=u(x,-p)$ is avoided in that case since the wavefunctions are complex).

\subsection{EE from the relativistic approximation for fermions}\label{app:relative-approx}

The formula \eq{SS-hardbox} for the non-relativistic fermions can be obtained alternatively from the well-known observation ({\it cf.} \eq{fermi-level-dispersion}, \eq{fermi-vel}) that fermions close to the Fermi surface have a relativistic dispersion relation. In fact, by introducing the 'particle' and 'hole' operators $\psi_{N-1+m}= p_m, m=1,2,..,\infty$, $\psi_{N-m}= h^\dagger_m, m=1,2,...,N$, the mode expansion \eq{second-q} becomes similar to that of a Dirac fermion with the particle and holes identified with particles and antiparticles respectively, with the important difference that the `antiparticle modes' have a natural UV cut-off given by the finite depth of the Fermi sea; in case of the hard box, the upper bound on the mode number $m_{max}= N$ translates to a UV cutoff on the momentum $|p_{max}| = \Lambda = \hbar m_{max} \pi/L$ $ = \fr{\hbar N\pi}{L}= p_F$. For the hard box, note that the ground state EE described in Appendix \ref{app:thomas-fermi}, involves the same wavefunctions $\chi_n(x) \propto \sin(n \pi x/L)$ (which are the same for the relativistic and for the non-relativistic problem--- the energy eigenvalues of course are different; but for the ground state EE they do not come into play). Hence ground state EE of Appendix \ref{app:thomas-fermi} can be interpreted as that of two species of free massless relativistic relativistic fermions with a cutoff $\Lambda = p_F $ \footnote{We can impose the same cut-off $\Lambda$ for the particle and antiparticle modes, as would be customary for a relativistic fermion. Since the particle modes do not contribute to the expressions e.g. \eq{inverse-droplet}, this is somewhat moot.}

Since two species of massless relativistic fermions comprise a conformal field theory of central charge $c=2 \times 1/2 =1$, the formula \eq{SS-hardbox} can be derived from the standard CFT formula for the ground state EE \cite{Holzhey:1994we, Calabrese:2004eu}.

\section{Lattice formulation of free relativistic scalars}\label{app:herzog}

In this section, we will review the standard lattice formulation \footnote{See, e.g. \cite{Creutz:1983njd}, Chapter 4. For lattice field theory references in the context of entanglement entropy, see, e.g. \cite{Peschel:2002yqj, Eisler:2009vye}. Note however that in our exact lattice boson, the role of the speed of light is played by the speed of sound $v_F$.} of a massless free scalar field theory in one dimension and its entanglement entropy.

\subsection{Lattice bosons in a hard box}\label{app:lattice-box}

Let us consider massless bosons in a hard box of length $L$, with boundary conditions $\phi=0$ at both ends. We will use $j=0,1,...,M, M+1$ lattice points, with lattice spacing $\epsilon = L/(M+1)$, with b.c. $\phi_0=0= \phi_{M+1}$. The independent lattice variables are (for $j=1,2,...,M$)  
\begin{align}
  \phi_j(t) & \equiv \phi(x_j) = \sum_{n=1}^{M} \frac{1}{\sqrt{2 \omega_n}}\sqrt{\frac{2}{L}} \sin(\frac{n \pi j}{M+1}) \left[a_n \exp(-i \omega_n t) + a^{\dagger}_n \exp(i \om_n t)\right] \label{phi-def-hard-app} \\
	\pi_j(t) &\equiv \pi(x_j) \epsilon = \sum_{n=1}^{N} i\epsilon \sqrt{\frac{\omega_n}{2}}\sqrt{\frac{2}{L}} \sin(\frac{n \pi j}{M+1}) \left[-a_n \exp(-i \om_n t) + a^{\dagger}_n \exp(i \om_n t)\right], \label{pi-def-hard-app}
\end{align}
The lattice Lagrangian and Hamiltonian are given by \eq{lattice-lag} and \eq{lattice-ham}
\begin{align}
 & {\cal L}_{lattice}= \frac{\epsilon}{2} \sum_j \left[ \dot{\phi_j}^2 -  v_F^2\frac{(\phi_{j+1} - \phi_j)^2}{\epsilon^2}  \right]
  \label{lattice-lag}\\
  & H_{lattice}= \frac{1}{2\epsilon} \sum_j\left[  \pi_j^2 + v_F^2 (\phi_{j+1} - \phi_j)^2 \right]
  \label{lattice-ham}
\end{align}
The equation of motion is the lattice Klein-Gordon equation
\begin{align}
  \frac{\ddot{\phi}_j}{v_F^2} = \frac{\phi_{j+1} + \phi_{j-1} - 2 \phi_j}{\epsilon^2}
  \label{lattice-KGE}
\end{align}
which leads to the following time-dependence
\begin{align}
  \omega_n = v_F \frac{2}{\epsilon}|\sin(\frac{n \pi}{2(M+1)})|
  \label{omega-n-app}
\end{align}
In terms of the oscillators, the Hamiltonian \eq{lattice-ham} becomes (dropping the zero point energy)
\begin{align}
  H_{lattice}= \sum_{n=1}^M \omega_n a^\dagger_n a_n
  \label{oscill-ham-app}
\end{align}

\subsection{Partition function from the lattice boson formulation}\label{app:lattice-partn-fn}

We have adopted the above lattice boson, with $M=N$, to construct the exact lattice bosonization described in Section \ref{sec:fixed-lat}. In this subsection we will compare thermodynamic properties of the lattice Hamiltonian with the exact oscillator Hamiltonian $H_1 + H_2$ of \eq{bose-expand}.


Let us compute the partition function $Z= \tr \exp[-\b H]$ for (a) $H= H_{lattice}$, (with $M=N$) (b) $H= H_1$, (c) $H_{total}= H_1 + H_2$ \footnote{In defining the partition function, we omit the zero-point energy $E_g$ in \eq{bose-expand}.} where $H_1, H_2$ are defined in \eq{bose-expand}. The Hamiltonians in cases (a) and (b) are linear in the oscillator number, hence the corresponding partition functions $Z_a$ and $Z_b$ are given by the standard formulae:
\begin{align}
   \log Z_a= -\sum_{k=1}^N \log(1- \exp[-\b \omega_k]),\kern30pt 
   \log Z_b= -\sum_{k=1}^N \log(1- \exp[-\b \a N k])
  \label{za-zb}
\end{align}
Here $\a = \hbar^2\pi^2/L^2$ (see \eq{hard-box}). Note that $\omega_k \approx \a N k$ for small $k$; hence at small enough temperatures (when effectively only the low $k$ modes are excited), we expect $Z_a$ and $Z_b$ to agree. Furthermore, under such conditions, i.e. in states where only low $k$ modes are excited, $H_2$ is subdominant compared to $H_1$, as noted below \eq{bose-expand}. Hence at low enough temperatures, we expect all three quantities $Z_a, Z_b$ and $Z_c$ to agree.

\begin{figure}[H]
\begin{center}
  \includegraphics[scale=0.5]{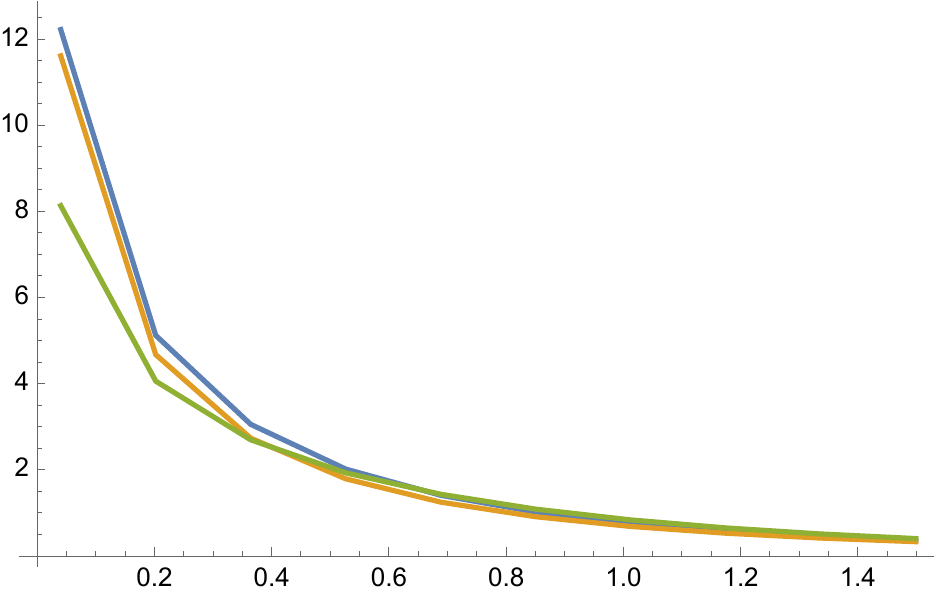}
\end{center}
\caption{\footnotesize Here the horizontal axis denotes $\b$, the inverse temperature, and the vertical axis the logarithm of the partition function. The blue line represents $\log Z_a$, the orange one $\log Z_b$ and the green one $\log Z_c$. The first two are calculated according to \eq{za-zb}, and the third one according to \eq{zc}. The parameter values are taken to be $mL^2 =2$, $\hbar=1/N$, $N=5$. For numerical calculation of $\log Z_c$, the sum over each occupation number $r_i$ in \eq{zc} is cut off at a maximum value $r_{max}=10$. As we see, the curves start merging when $\b \gg mL^2/\pi^2 \approx 0.2 $ (this inequality is explained in \eq{rel}).}
\label{fig:partn-fn}
\end{figure}

To verify these claims explicitly, and compare the three partition functions, we need to also compute $Z_c$; for this, we need to work a little harder, since $H_2$ contains terms non-linear in the occupation numbers, because of which formulas like \eq{za-zb} are not available. The problem is simplified, however, due to the integrable structure of the Hamiltonian (see Section \ref{sec:integrable}). Since the occupation number eigenstates $|r_1, r_2, ..., r_N \ran$ \eq{identity} are an eigenbasis of $H_1$ and $H_2$. It is straightforward to work out the eigenvalues:
\begin{align}
   E_1= \a N \sum_{k=1}^N k r_k, \kern30pt
   E_2= \fr\a{2}\left[ \left(- \sum_{k=1}^N \fr{k(k^2-1)}3 r_k \right) +
    \left(\sum_{k=1}^N \sum_{i=k}^N \sum_{j=k}^N r_i r_j \right)\right]
  \label{eigen-h1-h2}
\end{align}
The partition function is given simply by
\begin{align}
 \log Z_c= \log\left(\sum_{r_1=0}^\infty... \sum_{r_N=0}^\infty \exp[-\b (E_1 + E_2)]\right)
  \label{zc}
\end{align}
We present, in Figure \ref{fig:partn-fn}, results from a numerical calculation of \eq{za-zb} and \eq{zc}. We find that the three partition functions agree as long as $\b \gsim 1/\a N$. In the $N \to \infty$ limit, the restriction on temperature is removed; hence the lattice partition function $Z_a$ reproduces the full bosonic partition function $Z_c$.

\section{Periodic boundary condition}\label{app:periodic}

As emphasized before, we have mainly focussed in this paper on fermions in a hard box or confining potentials of a more general type. The single-particle eigenfunctions are therefore given by standing waves rather than travelling waves going left- or right (see Appendix \ref{app:p>0}). In this Appendix, we include a discussion of fermions on a circle (periodic one-dimensional box)

\subsection{Fermions}

For fermions in a circle of length $L$, the spectrum is given by
\begin{align}
  \chi_n(\l)= \sqrt{\fr1{L}} \exp(i\fr{2\pi n \l}{L}), \;
  \ep(n)=2 \a n^2,\; \a= \fr{\hbar^2 \pi^2}{L^2}, \quad n=0,\pm 1,\pm 2,...,\infty
\label{periodic-box}
\end{align}
Since the $\pm n$ eigenmodes are degenerate (corresponding to left- and right- moving modes), in order to arrange the spectrum according to the ordering (\ref{spectrum},\ref{ordering}), one can split the degeneracy by, say, raising the right-moving energies by an infinitesimal amount. This has been carried out in equations (19-21) of \cite{Dhar:2006ru}. The bosonized Hamiltonian has the same structure as in \eq{bose-expand}. The ground state energy can be easily worked out:
\begin{align}
  E_g & = \fr\a6 \left( N^3 - N \right), \; \hbox{if $N$ is odd}, \nonumber\\
  &= \fr\a6 \left( N^3 + 2N \right), \; \hbox{if $N$ is even}
  \label{exact-odd-even}
\end{align}
Note that the leading large $N$ behaviour is the same as in case of the hard box, as one would expect.

The EE in the periodic case, of an interval $[x_1, x_2]$ in the fermion theory, is again given by \eq{SS-v0}:
\begin{align}
  S_A= \frac13 \log(N \fr{l}{L})
  \label{SS-v0-circle}
\end{align}

\subsection{Lattice boson in the periodic case}

The structure of the bosonic Hamiltonian in this case is similar to \eq{bose-expand}, with a linear term $H_1$ describing low energy excitations above the ground state energy. Since $H_1$ has a linear spectrum, and the geometry of the fermion problem is circular, it is appropriate to define our lattice boson in terms of the lattice version of a massless relativistic boson on a circle. We accordingly consider a periodic lattice of size $L$, with lattice spacing $\epsilon= L/N$ (i.e. there are $N$ lattice points $x_j = j \epsilon$, $j=1,2,...,N$). The lattice boson is described by a set of $N$ values of the field $\phi(x_j)= \phi_j$, $j=1,2, ...,N$, and of the conjugate variable $\pi(x_j)= \pi_j/\ep$. The normal mode expansion on the lattice (obtained from a solution of the lattice Klein-Gordon equation \eq{lattice-KGE} with circular boundary conditions) are given by 
\begin{align}
  \phi_j(t) &= \sum_{n=-(N-1)/2}^{(N-1)/2} \frac{1}{\sqrt{2 \omega_n}} \left[\alpha_n \exp(i \frac{2\pi n j}{N}- i \om_n t) + \alpha^{\dagger}_n \exp(-i \frac{2 \pi n j}{N} + i \om_n t)\right] \label{phi-def-app} \\
	\pi_j(t) &= \sum_{n=-(N-1)/2}^{(N-1)/2} i \epsilon \sqrt{\frac{\omega_n}{2}} \left[-\alpha_n \exp(i \frac{2\pi n j}{N}- i \om_n t) + \alpha^{\dagger}_n \exp(-i \frac{2 \pi n j}{N}+ i \om_n t)\right], \label{pi-def-app}
\end{align}
We map the $\alpha_n$ operators to our exact boson operators $a_n$ in the following way. 
\begin{align}
\alpha_n = \begin{cases}
a_{2n+1} & n>0 \\
a_{2|n|} & n<0 \\
a_1 & n=0
\end{cases}
\end{align}
Of course, EE is not affected by this arbitrary choice of mapping. 
Here, the frequency (working in $v_F=1$ units)
\begin{align}
  \omega_n \equiv \sqrt{m^2 +  \fr4{\epsilon^2} \, \sin^2(\pi n/N)}
  \label{omega-n}
\end{align}
follows from the lattice Klein-Gordon equation with a mass term
\begin{align}
\ddot{\phi_j} = \frac{\phi_{j+1} + \phi_{j-1} - 2 \phi_j}{\epsilon^2} + m^2 \phi_j
\end{align} 
The mass $m$ is an infrared regulator that is being put in to handle the $n=0$ mode. In the quantum theory, the normal modes become annihilation and creation operators satisfying the Heisenberg algebra
\begin{align}
  [\alpha_k, \alpha^\dagger_l]= \delta_{kl}
  \label{lattice-osc}
\end{align}
which are equivalent to the equal-time commutation relations
\begin{align}
  [\phi_j, \pi_l]= i \delta_{kl}
  \label{lattice-heisen}
\end{align}
Again, the exact hamiltonian \eq{bose-ham} appropriate to the periodic boundary conditions, can be transcribed into a lattice Hamiltonian by using  the inverse relations:
\begin{align}
  & \alpha_n = \frac{1}{2N}\sum_j\left[\sqrt{2\om_n} \phi_j  + \frac{i}{\epsilon} \sqrt{\frac{2}{\om_n}} \pi_j \right] \exp(- i\frac{2\pi n j}{N})
  \nonumber\\
  & \alpha^{\dagger}_n = \frac{1}{2N}\sum_j\left[\sqrt{2\om_n} \phi_j  - \frac{i}{\epsilon} \sqrt{\frac{2}{\om_n}} \pi_j \right] \exp( i\frac{2\pi n j}{N})
  \label{inverse-a-ad}
\end{align}
The low energy hamiltonian will again be given by \eq{lattice-ham}.

\paragraph{Entanglement entropy}

For the periodic lattice it was found in \cite{Eisler:2009vye} \footnote{And derived analytically in \cite{Holzhey:1994we, Calabrese:2004eu, Casini:2005zv}} that
\begin{equation} \label{bose-EE-circular-lattice}
  S_A = \frac{1}{3} \log(\nu) =\frac{1}{3} \log(\fr{l}\ep) =\frac{1}{3} \log(\fr{l}{L/N})  = \frac13 \log(N \fr{l}{L})
\end{equation}
where $\nu= l/\ep$ is the number of lattice points in the region of interest $A$ of length $l$. We have independently verified this extensively numerically, using \eq{casini-huerta}.

Note that this formula reproduces the fermionic result \eq{SS-v0-circle}.

\section{Locality}\label{app:locality}

We attempt to map local fermionic operators to local bosonic operators for low-lying states. 

Say we consider the four lowest level states in the fermionic theory: $\ket{0}$, $a_1^{\dagger} \ket{0}$, $\frac{1}{\sqrt{2}} {a_1^{\dagger}}^2 \ket{0}$ and $a_2^{\dagger} \ket{0}$. The matrix elements of the fermionic operator $O_f = (\psi^{\dagger}(x_1)\psi(x_2) + \psi^{\dagger}(x_2)\psi(x_1))/2$ (so as to make it Hermitian) in the large $N$ approximation, where $x_1 = x + \eta$ and $x_2 = x - \eta$, is given by

\begin{blockarray}{ccccc}
& $\ket{0}$ & $a_1^{\dagger} \ket{0}$ & $\displaystyle \frac{1}{\sqrt{2}} {a_1^{\dagger}}^2 \ket{0}$ & $a_2^{\dagger} \ket{0}$  \\
\begin{block}{c(cccc)}
$\bra{0}$ & $k(\eta) \equiv \displaystyle \sum_{m=0}^{N-1}  \cos(m \eta)$ & $\cos(N\eta) e^{ix}$ & $\cos(N\eta) e^{i2x}$ & $\cos(N\eta) e^{i2x}$ \\
$\bra{0} a_1$ & $\cos(N\eta) e^{-ix}$ & $k(\eta)$ & $\cos(N\eta) e^{ix}$ & $\cos(N\eta) e^{ix}$  \\
$\bra{0} a_1^2 \displaystyle \frac{1}{\sqrt{2}}$ & $\cos(N\eta) e^{-i2x}$ & $\cos(N\eta) e^{-ix}$ & $k(\eta)$ & $0$ \\
$\bra{0} a_2$ & $\cos(N\eta) e^{-i2x}$ & $\cos(N\eta) e^{-ix}$ & $0$ & $k(\eta)$ \\
\end{block}
\end{blockarray}

We can try mapping the above fermion operator to the following bosonic operator:
\begin{multline}
  O_b = k(\eta)\mathds{1} + \cos(N\eta)[u_1 \;\phi(x) + u_2 \;\pi(x) + u_3 \; \partial_x \phi(x) + u_4 \; \partial_x \pi(x) + u_5 : \phi^2(x) : + \\ u_6 :\pi^2(x): + u_7 :\phi(x)\partial_x\pi(x): + u_8 :\partial_x \phi(x) \pi(x): + u_9 :\phi^3(x): + u_{10} :\pi^2(x) \phi(x):
    \label{local-map}
\end{multline}

The coefficients take the values given below.
\begin{align}
u_1 &= \frac{\sqrt{2} \omega_1 \sqrt{\omega_2} - 2\sqrt{2} \sqrt{\omega_1} \omega_2}{\omega_1 - 2 \omega_2} \nonumber \\
u_2 &= 0 \nonumber \\
u_3 &= 0 \nonumber \\
u_4 &= \frac{\sqrt{2}(\sqrt{\omega_1} - \sqrt{\omega_2})}{\omega_1 - 2 \omega_2} \nonumber \\
u_5 &= \frac{(\omega_1 \omega_2)^{3/2} \left(-\sqrt{2} \sqrt{\frac{\omega_1}{\omega_2}}+\sqrt{2}
   \sqrt{\frac{\omega_2}{\omega_1}}-4\right)}{2 \left(\omega_1^2-3 \omega_1 \omega_2+2 \omega_2^2\right)} \nonumber \\
u_6 &= -\frac{4 \sqrt{\omega_1
   \omega_2}+\sqrt{2} \omega_1-\sqrt{2} \omega_2}{4 \omega_1^2-6 \omega_1 \omega_2+2 \omega_2^2} \nonumber \\
u_7 &= \frac{2 \sqrt{2} \omega_1^3-5 \sqrt{2} \omega_1^2 \omega_2-4
   \sqrt{\omega_1 \omega_2^5}+4 \sqrt{2} \omega_1 \omega_2^2+8 (\omega_1 \omega_2)^{3/2}-\sqrt{2} \omega_2^3}{2 \left(2 \omega_1^3-7 \omega_1^2 \omega_2+7 \omega_1
   \omega_2^2-2 \omega_2^3\right)} \nonumber \\
u_8 &= \frac{-4 \sqrt{\omega_1^5 \omega_2}+\sqrt{2} \omega_1^3-4 \sqrt{2} \omega_1^2 \omega_2+5 \sqrt{2} \omega_1
   \omega_2^2+8 (\omega_1 \omega_2)^{3/2}-2 \sqrt{2} \omega_2^3}{2 \left(2 \omega_1^3-7 \omega_1^2 \omega_2+7 \omega_1 \omega_2^2-2 \omega_2^3\right)} \nonumber \\
u_9 &= -\frac{4 \left(\sqrt{2}-1\right) \omega_1^{3/2} (\omega_1-2
   \omega_2)}{\left(6+\sqrt{2}\right) \omega_1-12 \omega_2} \nonumber \\
u_{10} &= \frac{2 \left(\sqrt{2}-2\right)
   \sqrt{\omega_1}}{\left(6+\sqrt{2}\right) \omega_1-12 \omega_2}
\end{align}

\section{Justification for the coordinate transformation}\label{app:coord}

In this section, we provide a justification for the coordinate transformation given in \eqref{non-uniform}.

In brief, the logic of the coordinate transformation is as follows: in the fermion theory a change in the single particle hamiltonian \eq{single-ham}, in particular a change in the potential $V(x)$, can be implemented by a canonical transformation in the single particle phase space.\footnote{In the quantum theory, the canonical transformation becomes a $W_\infty$ transformation \cite{Dhar:1992hr, Dhar:1992rs, Das:1991uta}.} Such a canonical transformation, when applied to a droplet which represents the ground state of the original Hamiltonian, generates a deformed droplet which is the ground state of the deformed Hamiltonian; when projected on to the $x$-axis, the deformation of the boundary of the droplet, defines a transformation on the $x$-axis. Stated in another way, the canonical transformation  ($x,p \to \tilde x, \tilde p$), when restricted to the Fermi surface, becomes a coordinate transformation $(x \to \tilde x$). It is this coordinate transformation which we use in \eq{non-uniform}. The coordinate transformation, therefore, is a direct consequence of the change in the potential. 

Let us elaborate on the above. Let us start with a single-particle hamiltonian in the fermion theory of the form
\begin{align}
  h(x,p) = \frac{p^2}{2} + V(x) 
  \label{single-ham-app}
\end{align}
In the large $N$ limit (semiclassical limit), the Wigner phase space density takes the droplet form
\begin{align}
  u(x,p) = \Theta( \epsilon_F - h(x,p))
  \label{droplet-app}
\end{align}
as seen previously in \eq{droplet}; here $\ep_F$ is defined by the normalization condition $\int \fr{dx\,dp}{2\pi} u(x,p)= \hbar N = 1$.

To implement a change in the potential $V(x)$, we consider a canonical transformation defined by the function
\begin{align}
  W(x,p) = -p a(x)
  \label{special-w}
\end{align}
which generates the following symplectic flow (Hamiltonian flow) in phase space (where $s$ is the transformation parameter)
\begin{align}
\frac{dx}{ds} &= \frac{\partial W}{\partial p} = - a(x) \nonumber \\
\frac{dp}{ds} &= - \frac{\partial W}{\partial x} = p a'(x)
\label{flow}
\end{align}
This ``motion'' can be integrated to give
\begin{align}
(x_0, p_0) \to (x_s(x_0), p_s(x_0,p_0)): \kern20pt 
& \alpha(x_s) = -s + \alpha(x_0), \kern10pt \alpha'(x) = 1/a(x) \\
  & p_s = p_0 \frac{a(x_0)}{a(x_s)}
  \label{x-p-trans}
\end{align}
where $\alpha'(x) = 1/a(x)$. Note that due to the special choice of the canonical transformation \eq{special-w}, the transformation of $x$ does not involve $p$.

What happens to the Hamiltonian? We find
\begin{align}
  h(x_0, p_0) &= \frac{p_0^2}{2} + V(x_0)= \left(\frac{a(x_s)}{a(x_0)}\right)^2 \fr{p_s^2}{2} + V(x_0(x_s))  \nonumber\\
  & = \left(\frac{a(x_s)}{a(x_0)}\right)^2 \tilde h(x_s, p_s), \quad
  \tilde h(x_s, p_s)= \frac{p_s^2}{2} + \tilde V(x_s), \nonumber\\
  \tilde V(x_s) &= V(x_0(x_s))\left(\frac{a(x_0)}{a(x_s)}\right)^2
  \label{gen-ham}
  \end{align}

How does the ground state droplet \eq{droplet-app} transform? The initial droplet transforms as follows:
\begin{align}
  u_0(x_0, p_0) &= \Theta\left(\ep_{F,0}- (\frac{p_0^2}{2} + V(x_0))\right) =
  \Theta\left(\tilde\ep_F -(\frac{p_s^2}{2}
  + \tilde V(x_s))\right), \nonumber\\
  \tilde \ep_F & \equiv  \ep_{F,0}\left(\frac{a(x_0)}{a(x_s)}\right)^2
  \label{droplet-trans}
\end{align}
In the second step we have used the scale-invariance of the theta-function: $\Theta(\a x)= \Theta(x)$, $\a >0$. In general, this $\tilde{\epsilon}_F$ is position-dependent and is therefore not the new Fermi energy. We get the new potential $V_s(x)$ by adding the position-dependent part of $\tilde{\epsilon}_F$ together with $\tilde{V}$. The leftover constant piece of $\tilde{\epsilon}_F$, which we will call $\epsilon_{F,s}$, becomes the new Fermi energy. If $\tilde{\epsilon}_F$ has no position dependence, then $V_s = \tilde{V}$. The initial Fermi surface $p_0= p_{F,0}(x_0)$ is obtained by vanishing of the argument of the first theta-function, leading to
\[
p_{F,0}(x_0)= \sqrt{2(\ep_F - V(x_s))}.
\]
The final Fermi surface $p_s= p_{F,s}(x_s)$ is obtained by vanishing of the argument of the second theta-function, leading to
\[
p_{F,s}(x_s)= \sqrt{2(\tilde\ep_F - \tilde V(x_s))}.
\]
By using the transformation properties of the Fermi energy and the potential \eq{gen-ham} and \eq{droplet-trans}, it is easy to see that
\begin{align}
  \fr{p_{F,s}(x_s)}{p_{F,0}(x_0)}= \frac{a(x_0)}{a(x_s)}
  \label{pf-ratio}
\end{align}
Let us consider a pair of nearby points $(x_0, x_0+ \delta x_0)$ and their transforms $(x_s, x_s + \delta x_s)$, then from \eq{x-p-trans}, we see that
\begin{align}
\a'(x_0) \delta x_0=  \a'(x_s) \delta x_s \Rightarrow
\fr{\delta x_s}{\delta x_0}= \frac{a(x_s)}{a(x_s)}
\label{del-x-ratio}
\end{align}
where we have used $\a'(x)=1/a(x)$. 
Using \eq{pf-ratio} and \eq{del-x-ratio} we get
\begin{align}
p_{F,0}(x_0) \delta x_0= p_{F,s}(x_s) \delta x_s 
  \label{invariance-w}
\end{align}
Thus, the canonical transformation \eq{x-p-trans}, projected on to the Fermi surface (defined above by the boundary of the liquid droplet or by the vanishing of the argument of the theta-function), reduces to the coordinate transformation \eq{invariance} used in defining the lattice boson in a general confining potential.

\paragraph{Geometrical derivation} It is easy to derive the above fact by a simple geometrical observation. In Figure \ref{fig:potential} the droplet in panel (c) is transformed under the canonical transformation to the droplet in panel (d). Suppose that the points $(x_1, x_2)$ are transformed to the points $(\tilde x_1, \tilde x_2)$; then the following are true:\\
(i) The vertical lines $(x=x_1, x=x_2)$ transform to the vertical lines $(\tilde x=\tilde x_1, \tilde x= \tilde x_2)$ (this is because, due to the special nature of the canonical transformations \eq{x-p-trans}, the $x$-transformation is independent of $p$).\\
(ii) By construction, the boundary of the droplet transforms to the boundary of the droplet, and a droplet that is initially symmetric with respect to the x axis continues to stay symmetric with respect to the x axis under this transformation.\\
Thus, the blue strip flanked by the two vertical lines and ending at the boundary of the droplet in panel (c) must transform into the similar blue strip in panel (d). Since we are discussing here a canonical transformation, the area of the two blue strips must be the same, which implies
\begin{align}
2 \int_{x_1}^{x_2} d x\,  p_F( x) =
2 \int_{\tilde x_1}^{\tilde x_2} d\tilde x\, \tilde p_F(\tilde x)
\label{geometric}
\end{align}
For small enough intervals $[x_1, x_2]$ in which the variation of $p_F(x)$ can be ignored, we get \eq{invariance-w}.  [QED]

In the following, we elaborate this general procedure in two examples.

\subsection{Example 1: generating constant scaling of a hard box}

Let us take $a(x)= a_1 x$ in \eq{special-w}. This generates the scaling transformation
\begin{align}
  x_s= x_0 e^{-a_1 s},\; p_s = p_0 e^{a_1 s} \nonumber
\end{align}
The hard box potential can be represented as
\begin{align}
  V_L(x) := \bar{V} \Theta\left(x^2 -\frac{L^2}{4} \right)
  \label{hard-box-app}
\end{align}
where we are supposed take the infinite barrier limit $\bar V \to \infty$. The equation \eq{gen-ham} now reads as
\begin{align}
  h(x_0, p_0) &= \frac{p_0^2}{2} + \bar{V} \Theta\left(x_0^2 -\frac{L^2}{4} \right)  \nonumber\\
  & = e^{-2 a_1 s} \tilde h(x_s, p_s), \quad
  \tilde h(x_s, p_s)= \frac{p_s^2}{2} + \tilde V(x_s), \nonumber\\
  \tilde V(x_s) &=e^{2 a_1 s} \bar{V} \Theta\left(x_s^2 -\frac{L_s^2}{4} \right), \quad L_s= e^{-a_1 s}L_0
  \label{hard-box-ham}
  \end{align}
Note that we are interested in the limit $\bar V \to \infty$, hence the new potential ($V_s = \tilde{V}$ since $\tilde{\epsilon}_F$ is position independent in \eqref{new-ep-f}) simply represents a hard box of a scaled length $L_s$.

How does the semiclassical droplet transform that represents the ground state of the $N$ fermion system?
\begin{align}
  u_0(x_0, p_0) &= \Theta(\ep_{F,0} - \fr{p_0^2}{2})\Theta\left(x_0^2 -\frac{L^2}{4} \right) \nonumber\\
  &= \Theta(\ep_{F,0} - e^{-2 a_1 s} \fr{p_s^2}{2}) \Theta\left(x_0^2 -\frac{L^2}{4} \right)= \Theta(\ep_{F,s} - \fr{p_s^2}{2})\Theta\left(x_s^2 -\frac{L_s^2}{4} \right) \equiv u_s(x_s,p_s) \nonumber
\end{align}
where
\begin{align}
  \ep_{F,s}= e^{2 a_1 s} \ep_{F,0}
  \label{new-ep-f}
\end{align}
Note that the scaling of the Fermi energy is consistent with the fact that
$\ep_F =  N^2 \pi^2 \hbar^2/(2 L^2)$ and that $L_s \propto \exp[-a_1 s]$. Note that we have used here the transformation property that the phase space density $u(x,p)$ behaves like a scalar under a canonical transformation since the phase space measure remains invariant.

It is trivial to check that the coordinate transformation $x_0 \to x_s$ satisfies the defining property \eq{invariance}:
\[
\sqrt{\ep_{F,s}} \Delta x_s = \sqrt{\ep_{F,0}} \Delta x_0
\]
  
\subsection{Example 2: generating harmonic potential from a hard-box potential}\label{app:gen-harmonic}

We write below a generating function that takes us from a hard box potential (at $s=0$) to a harmonic potential (cut-off by a box, see Fig. \ref{fig:potential}).

\begin{figure}[H]
\begin{center}
\includegraphics[scale=.45]{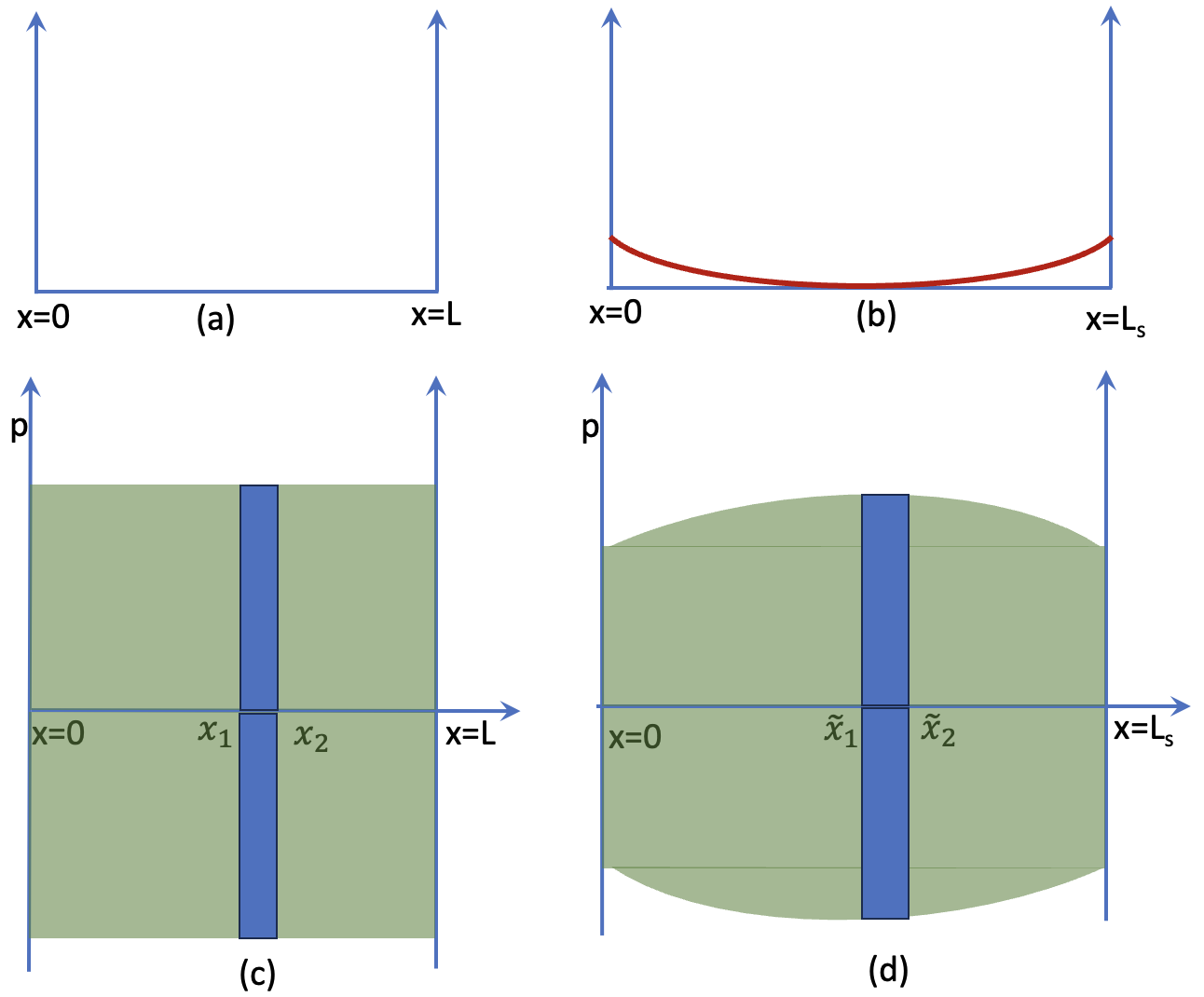}
\end{center}
\caption{\footnotesize This figure explains how a canonical transformation described by \eq{x-p-trans} can transform a hard box potential (panel (a), \eq{hard-box-app}) to a (truncated) harmonic potential (panel(b), \eq{cubic-v-ef}). The corresponding Fermi ground state droplets are depicted in panels (c) and (d), respectively. It is shown in remarks above \eq{geometric} that the blue strip in (c) transforms to the blue strip in (d), which provides a geometric derivation of \eq{invariance-w}.}
\label{fig:potential}
\end{figure}

Let us consider a canonical transformation generated by
\begin{align}
  W(x,p) =  p (a_1 x + a_3 x^3)
  \label{cubic}
\end{align}
The coordinate and momentum transform as
\begin{align}
x_s &= \sqrt{\frac{a_1}{-a_3 + e^{-2a_1 s} (a_3 + a_1/x_0^2)}} \nonumber \\
p_s &= p_0 f_p(s,x_0),\; f_p(s,x_0)= \frac{x_0\sqrt{-a_3 + e^{-2 a_1 s}(a_3 + a_1/x_0^2)}(a_1 - (e^{2 a_1 s} -1)a_3 x_0^2)}{a_1^{3/2}}
\label{x-p-cubic}
\end{align}
We can invert these relations (by replacing $s$ with $-s$) to get
\begin{align}
x_0 &= \sqrt{\frac{a_1}{-a_3 + e^{2a_1 s} (a_3 + a_1/x_s^2)}} \nonumber \\
p_0 &= p_s  \frac{x_s\sqrt{-a_3 + e^{2 a_1 s}(a_3 + a_1/x_s^2)}(a_1 - (e^{-2 a_1 s} -1)a_3 x_s^2)}{a_1^{3/2}}
\label{cubic-invert}
\end{align}
Following the steps outlined above, one can again find the modified Hamiltonian $\tilde h(x,p)$, the potential $V_s(x)$, and the Fermi surface $\ep_{F,s}$. For small $a_3$, their expressions are as follows:
\begin{align}
  \tilde h(x_s, p_s) &= \fr{p_s^2}{2} + V_s(x_s) \nonumber\\
 V_s(x) &=  \left(1-  \frac{3 a_{3,s}}{a_1} x^2\right)e^{-2 a_1 s} \bar{V}\Theta\left(x^2 - \frac{L_s^2}{4}\right) + \frac{3 a_{3,s}}{a_1} \epsilon_{F,s} x^2 \nonumber \\
a_{3,s} &= a_3 (1-e^{-2 a_1 s} ) \nonumber \\
L_s &= L e^{a_1 s}\sqrt{1+ \frac{a_{3,s}}{4 a_1} (L e^{ a_1 s})^2}  \nonumber \\
\epsilon_{F,s} &= \epsilon_F e^{-2 a_1 s}
\label{cubic-v-ef}
\end{align}
This new potential looks like a SHO placed inside a modified hard box of length $L_s$ (see Figure \ref{fig:potential}). It is straightforward to explicitly check the invariance property \eq{invariance} in this case. Keeping terms only up to first order in $a_3$, in consistency with the approximation we used above, we find (for some interval $\delta x_s$ inside the harmonic oscillator hard box) 
\begin{align}
p_{F,s} \delta x_s &= \sqrt{\epsilon_{F,s} - V_s(x_s)} \delta x_s \nonumber \\
&= \sqrt{\epsilon_F e^{-2 a_1 s} - 3 a_3 \frac{(1-e^{-2a_1 s})}{a_1} \epsilon_F e^{-2 a_1 s} x_s^2} \left( \frac{a_1}{-a_3 + e^{-2 a_1 s} (a_3 + a_1/x_0^2)} \right)^{3/2} \frac{e^{-2 a_1 s}}{x_0^3} \delta x_0 \nonumber \\
&= \sqrt{\epsilon_F e^{-2 a_1 s} - 3 a_3 \frac{(1-e^{-2 a_1 s} )}{a_1} \epsilon_F x_0^2}\sqrt{\frac{e^{-4 a_1 s}}{x_0^6} \left(e^{6 a_1 s} x_0^6 + \frac{3 a_1}{e^{-4a_1 s}a_1^2} e^{4 a_1 s} x_0^8 (1-e^{-2 a_1 s}) a_3\right)} \delta x_0 \nonumber \\
&= \sqrt{\epsilon_F} \;\delta x_0 \nonumber \\
&= p_F(x_0) \delta x_0
\label{check-invariance}
\end{align}
which provides an explicit check of the invariance property stated in \eq{invariance-w}.

\subsection{Coordinate transformation for the bosonic field}

In the above, we have explained how the coordinate transformation used in \eq{non-uniform} arises in a natural way in the fermion theory. In particular, how it arises as a consequence of a $W_\infty$ transformation of the kind \eq{special-w}. In Section \ref{sec:non-uniform-lat}, we have, of course, applied this transformation to the bosonic field. Can we have a direct justification for that in the bosonic theory itself?

Note that the $W_\infty$ transformation, in the quantum theory, can be regarded as a unitary operation on the second quantized fermion theory. Since the Hilbert space of this theory is isomorphic to the Hilbert space of the second quantized bosonic theory through the map described in Section \ref{sec:exact}, the above-mentioned $W_\infty$ transformation naturally acts on the bosonic states and bosonic fields.

Since the $W_\infty$ transformations used above are restricted to near the Fermi surface, in the bosonic theory we should look at this transformation on low energy bosonic states. In Appendix \ref{app:locality}, it has been shown that on such states, local fermion fields are simply related to local bosonic fields (see \eq{local-map}); at the lowest order
\begin{align}
 O_f= \fr12 \left(\psi^{\dagger}(x_1)\psi(x_2) + \psi^{\dagger}(x_2)\psi(x_1)\right)
  = O_b= k(\eta)\mathds{1} + \cos(N\eta)[u_1 \;\phi(X) + u_2 \;\pi(X) + \ldots
    \label{local-map-lowest}
\end{align}
where $x_1 = X + \eta$ and $x_2 = X - \eta$. An infinitesimal version of the $W$-transformation \eq{flow} on the $x_1, x_2$ reads\footnote{The transformation \eq{flow} acted on phase space, but evaluated near the Fermi surface it boils down to just a transformation on $x$; as mentioned above, this restriction is taken care of in the bosonic theory by restricting to low energy states.}:
\begin{align}
  & \delta x_1 - a(x_1), \;  \delta x_2 =  - a(x_2)
  \label{fermion-coord}
\end{align}
Assume $a(x)$ is a slowly varying function of $x$, and $x_1$ is close enough to $x_2$ so that $a(x_1) \approx a(x_2) \approx a(X)$. In this case, the transformation becomes
\begin{align}
  & \delta X =  - a(X), \; \delta \eta =0  
  \nonumber
\end{align}
Note that under these transformations
\begin{align}
& \delta(\psi^\dagger(x_1) \psi(x_2)) =  - a(X) \del_{x_1} \psi^\dagger(x_1)\psi(x_2) - a(X)\psi^\dagger(x_1) \del_{x_2} \psi^\dagger(x_2)) \nonumber\\
  & = - a(X)(\del_{x_1} + \del_{x_2})(\psi^\dagger(x_1) \psi(x_2)) \nonumber\\
  & = - a(X)\del_X (\psi^\dagger(x_1) \psi(x_2)) \nonumber
  \nonumber
\end{align}
It is easy, then, to see the following
\begin{align}
\delta O_f= -a(X) \del_X O_f
  = -a(X) \del_X O_b= \cos(N\eta)[- a(X) \del_X (u_1 \phi(X) + u_2 \pi(X))] + \ldots
    \label{local-deformation}
\end{align}
which goes to show that the local coordinate transformations \eq{fermion-coord} induce an identical coordinate transformation on the bosonic fields:
\begin{align}
\delta \phi(X)= - a(X) \del_X \phi(X), \;  \delta \pi(X)= - a(X) \del_X \pi(X),
    \label{boson-coord}
\end{align}
The latticized version of these transformations corresponds precisely to the coordinate transformations \eq{non-uniform}, \eq{phi-def-nonuniform}, \eq{pi-def-nonuniform}.


\section{Schur polynomials and giant gravitons}\label{app:schur}

The mathematics preliminaries for this appendix can be found, e.g. in \cite{fulton1991representation}. The physics background can be found, e.g. in \cite{Balasubramanian:2001nh, Corley:2001zk, Berenstein:2004kk}.

Let us consider a state of multiple giant gravitons where $r_k$ giant gravitons have angular momentum $k$, $k=1,2, ...,N$. According to \cite{Balasubramanian:2001nh, Corley:2001zk, Berenstein:2004kk, Suryanarayana:2004ig} such a state is created by a Schur polynomial $\chi_R(Z)$, where $R$ is an irrep given by a Young tableau characterized by $r_k$ columns, each with $k$ boxes, $k=1,2, ...,N$. We will examine a few examples where such a Schur polynomial, acting on the ground state of the matrix QM (corresponding to the filled Fermi sea of $N$ fermions) produces an excited $N$-fermion state with filled levels $f_n$ related to the $r_k$ precisely by the map \eq{rk-fn}.

Example 1:

We will first consider a single giant graviton of angular momentum $n$. This corresponds to a Young tableau with $r_n=1$, and all other $r_k=0$, $k \ne n$. According to the map \eq{rk-fn}, this corresponds to the fermion levels $f_1=0, .., f_{N-n+1}=N-n, ..., f_N=N$, which corresponds to a single hole at depth $n$.

For simplicity, we will take $N=3$, $n=2$. We will also slightly modify the problem to the case of a hermitian matrix $M$ in a simple harmonic oscillator potential (the generalization to a complex matrix $Z$ is simple). We will show that the Schur polynomial $\chi_n(M)$, acting on the ground state, produces a hole at depth $n=2$. 

Thus, let us consider $N=3$ fermions in a simple harmonic oscillator potential, with single particle hamiltonian
\[
h= \hbar^2(-\frac12 \frac{\del^2}{\del\l^2}) + \frac{\l^2}{2}
\]
and eigenfunctions $\chi_n(\l): h \chi_n(\l)= (n+\frac12) \hbar$.
The ground state of this system is given by, up to normalization, by the Slater determinant:
\begin{align}
  \psi_0(M)=\left( \begin{smallmatrix}
  \chi_0(\l_1)  & \chi_0(\l_2) &  \chi_0(\l_3) \\
  \chi_1(\l_1)  & \chi_1(\l_2) &  \chi_1(\l_3) \\
  \chi_2(\l_1)  & \chi_2(\l_2) &  \chi_2(\l_3)
  \end{smallmatrix} \right)
  = {\cal N} \Delta(\l) \exp[-\frac12(\l_1^2 + \l_2^2 + \l_3^2)].
  \label{slater}
\end{align}
Here ${\cal N}$ is a normalization constant and we have used the notation 
\begin{align}
\Delta(\l)= 
   \left(
   \begin{smallmatrix}
  1  & 1 &  1 \\
  \l_1  & \l_2 &  \l_3 \\
  \l_1^2  & \l_2^2 &  \l_3^2
\end{smallmatrix}
   \right)
   \label{van-der}
\end{align}
for the Vandermonde determinant. Let us consider the giant graviton
operator corresponding to the Young tableaux
\raise3pt\hbox{\ytableausetup
 {mathmode, boxframe=normal, boxsize=.5em}
 \begin{ytableau}
   {}\\
   {}
\end{ytableau}}, namely
\[
\chi_{\ytableausetup
 {mathmode, boxframe=normal, boxsize=.3em}
 \begin{ytableau}
   {}\\
   {}
\end{ytableau}}(M)= \frac12 (\tr M)^2 - \frac12 \tr M^2
=\sum_{i<j=1}^3 \l_i\l_j 
\]
It is easy to show that
\[
\chi_{\ytableausetup
 {mathmode, boxframe=normal, boxsize=.3em}
 \begin{ytableau}
   {}\\
   {}
\end{ytableau}}(M) \Delta(\l)= \left(
   \begin{smallmatrix}
  1  & 1 &  1 \\
  \l_1^2  & \l_2^2 &  \l_3^2 \\
  \l_1^3  & \l_2^3 &  \l_3^3
\end{smallmatrix}
   \right)
   \]
   which can be reinterpreted as the following equation:
\[
\chi_{\ytableausetup
 {mathmode, boxframe=normal, boxsize=.3em}
 \begin{ytableau}
   {}\\
   {}
\end{ytableau}}(M) \left( \begin{smallmatrix}
  \chi_0(\l_1)  & \chi_0(\l_2) &  \chi_0(\l_3) \\
  \chi_1(\l_1)  & \chi_1(\l_2) &  \chi_1(\l_3) \\
  \chi_2(\l_1)  & \chi_2(\l_2) &  \chi_2(\l_3)
  \end{smallmatrix} \right) = \left( \begin{smallmatrix}
  \chi_0(\l_1)  & \chi_0(\l_2) &  \chi_0(\l_3) \\
  \chi_2(\l_1)  & \chi_2(\l_2) &  \chi_2(\l_3) \\
  \chi_3(\l_1)  & \chi_3(\l_2) &  \chi_3(\l_3)
  \end{smallmatrix} \right)
   \]   
 In the notation of Figure \ref{fig:young}, the above equation becomes
\begin{figure}[H]
\begin{center}
\ytableausetup
 {mathmode, boxframe=normal, boxsize=2em}
 \begin{ytableau}
   \none\\
 {} \\
 {}  
 \end{ytableau}
  \ytableausetup
 {mathmode, boxframe=normal, boxsize=1em}
\begin{ytableau}
  \none[\vdots] \\
  \scriptstyle 6 \\
  \scriptstyle 5\\
  \scriptstyle 4\\
  \scriptstyle 3\\
  *(red)\scriptstyle 2\\
  *(red) \scriptstyle 1\\
  *(red) \scriptstyle 0
\end{ytableau}
\kern20pt
\lower25pt\hbox{=}
\kern20pt
\begin{ytableau}
  \none[\vdots] \\
  \scriptstyle 6 \\
  \scriptstyle 5\\
  \scriptstyle 4\\
  *(red)\scriptstyle 3\\
  *(red)\scriptstyle 2\\
  \scriptstyle 1\\
  *(red) \scriptstyle 0
\end{ytableau}
	\caption{\footnotesize A single giant graviton of angular momentum $n=2$ creates a hole at depth $n$.}
	\label{fig:young-a}
\end{center}
\end{figure}   

Example 2:

For our next non-trivial example, let us consider two giant gravitons of different angular momenta, say $n_1$ and $n_2$ (say $n_2 \geq n_1$). The corresponding Young tableau is given by $r_{n_1} = 1 = r_{n_2}$ and all other $r_k=0$. The map \eqref{rk-fn} then tells us that this corresponds to the fermion state $f_1 = 0, \ldots, f_{N-n_2} = N - n_2 - 1, f_{N-n_2+1} = N-n_2 +1, \ldots, f_{N-n_1} = N - n_1, f_{N-n_1 +1} = N-n_1 +1, \ldots f_{N} = N+1$, i.e. two holes at depth $n_1$ and $n_2$. Once again, we can explicitly check that this is true for $N=3$ and $n_1 = 1, n_2=2$. The giant graviton operator corresponding to the Young tableau \raise3pt\hbox{\ytableausetup
 {mathmode, boxframe=normal, boxsize=.5em}
 \begin{ytableau}
   {} & {}\\
   {}
\end{ytableau}} is given by

\begin{align}
\chi_{\raise3pt\hbox{\ytableausetup
 {mathmode, boxframe=normal, boxsize=.5em}
 \begin{ytableau}
   {} & {}\\
   {}
\end{ytableau}}}(M) = \frac{1}{3} (\tr M)^3 - \frac{1}{3} \tr M^3
\end{align}
When acted on $\Delta(\lambda)$, it gives
\begin{align}
\chi_{\raise3pt\hbox{\ytableausetup
 {mathmode, boxframe=normal, boxsize=.5em}
 \begin{ytableau}
   {} & {}\\
   {}
\end{ytableau}}}(M) \Delta(\lambda) = \left(
   \begin{smallmatrix}
  1  & 1 &  1 \\
  \l_1^2  & \l_2^2 &  \l_3^2 \\
  \l_1^4  & \l_2^4 &  \l_3^4
\end{smallmatrix}
   \right),
\end{align}
which shows that
\begin{align}
\chi_{\ytableausetup
 {mathmode, boxframe=normal, boxsize=.3em}
 \begin{ytableau}
   {} & {}\\
   {}
\end{ytableau}}(M) \left( \begin{smallmatrix}
  \chi_0(\l_1)  & \chi_0(\l_2) &  \chi_0(\l_3) \\
  \chi_1(\l_1)  & \chi_1(\l_2) &  \chi_1(\l_3) \\
  \chi_2(\l_1)  & \chi_2(\l_2) &  \chi_2(\l_3)
  \end{smallmatrix} \right) = \left( \begin{smallmatrix}
  \chi_0(\l_1)  & \chi_0(\l_2) &  \chi_0(\l_3) \\
  \chi_2(\l_1)  & \chi_2(\l_2) &  \chi_2(\l_3) \\
  \chi_4(\l_1)  & \chi_4(\l_2) &  \chi_4(\l_3)
  \end{smallmatrix} \right)
\end{align}

\begin{figure}[H]
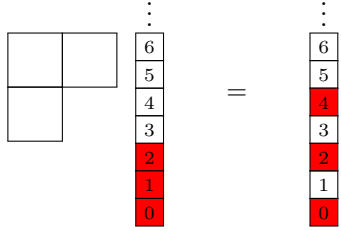

\begin{center}
\ytableausetup
 {mathmode, boxframe=normal, boxsize=2em}
 \begin{ytableau}
   \none\\
 {} & {} \\
 {}  
 \end{ytableau}
  \ytableausetup
 {mathmode, boxframe=normal, boxsize=1em}
\begin{ytableau}
  \none[\vdots] \\
  \scriptstyle 6 \\
  \scriptstyle 5\\
  \scriptstyle 4\\
  \scriptstyle 3\\
  *(red)\scriptstyle 2\\
  *(red) \scriptstyle 1\\
  *(red) \scriptstyle 0
\end{ytableau}
\kern20pt
\lower25pt\hbox{=}
\kern20pt
\begin{ytableau}
  \none[\vdots] \\
  \scriptstyle 6 \\
  \scriptstyle 5\\
  *(red)\scriptstyle 4\\
  \scriptstyle 3\\
  *(red)\scriptstyle 2\\
  \scriptstyle 1\\
  *(red) \scriptstyle 0
\end{ytableau}
	\caption{\footnotesize Two giant gravitons of angular momentum $n_1 = 1$ and $n_2 = 2$ create two holes at depth $1$ and $2 $.}
	\label{fig:young-b}
\end{center}
\end{figure}

\section{More details on bosonization}\label{app:boson}

In \eq{fermionize1}, \eq{fermionize2}, the bosonic oscillators are written in terms of the fermion bilinears $\psid_m \psi_n$. 
The inverse relation, which was also worked out in \cite{Dhar:2005fg}, is
\begin{align}
& \psid_{n+m}~\psi_n  \nonumber\\
& = \sum_{k=1}^{N-1} \biggl[\sigma_k^m~
{\sigma_{k+1}^\dagger}^m~\theta_+(\ad_k a_k-m) - \sum_{r_k=0}^\infty~
\sigma_{k-1}^{m-1-r_k}~{\sigma_k^\dagger}^{m-2-r_k}~\sigma_k^{r_k}~
{\sigma_{k+1}^\dagger}^{r_k+1}  \nonumber\\
& ~~~~~~~~~\times \theta_-(\ad_k a_k-m+1)~
\theta_+(\ad_{k-1} a_{k-1}+\ad_k a_k-m+1)~
\delta(\ad_k a_k-r_k)  \nonumber\\
& ~~~~~~+ \sum_{j=2}^{k-1} (-1)^j \sum_{r_{k-j+1}=0}^\infty 
\sum_{r_{k-j+2}=0}^\infty \cdots \sum_{r_k=0}^\infty 
\sigma_{k-j}^{m-j-\sum_{i=1}^j r_{k-j+i}}   \nonumber\\
& ~~~~~~~~~\times {\sigma_{k-j+1}^\dagger}^{m-j-1-\sum_{i=1}^j r_{k-j+i}}~
\sigma_{k-j+1}^{r_{k-j+1}}~
{\sigma_{k-j+2}^\dagger}^{r_{k-j+1}} \cdots \sigma_{k-1}^{r_{k-1}}~
{\sigma_k^\dagger}^{r_{k-1}}  \nonumber\\
& ~~~~~~~~~\times \sigma_k^{r_k}~{\sigma_{k+1}^\dagger}^{r_k+1}~
\theta_- \biggl (\sum_{i=1}^j\ad_{k-j+i}~a_{k-j+i}-m+j \biggr)  \nonumber\\
& ~~~~~~~~~\times 
\theta_+ \biggl (\sum_{i=0}^j \ad_{k-j+i}~a_{k-j+i}-m+j \biggr)~
\Pi_{i=1}^j \delta(\ad_{k-j+i}~a_{k-j+i}-r_{k-j+i})  \nonumber\\
& ~~~~~~+ (-1)^k \sum_{r_1=0}^\infty \cdots \sum_{r_k=0}^\infty 
{\sigma_1^\dagger}^{m-1-k-\sum_{i=1}^k r_i}~\sigma_1^{r_1}~
{\sigma_2^\dagger}^{r_1} \cdots \sigma_{k-1}^{r_{k-1}}~
{\sigma_k^\dagger}^{r_{k-1}} \nonumber\\
& ~~~~~~~~~\times \sigma_k^{r_k}~{\sigma_{k+1}^\dagger}^{r_k+1}~ 
\theta_- \biggl (\sum_{i=1}^k\ad_i a_i-m+k \biggr)~
\Pi_{i=1}^k \delta(\ad_i a_i-r_i) \biggr ] \nonumber\\
& ~~~~~~~~~\times
\delta \biggl (\sum_{i=k+1}^N \ad_i a_i-n+N-k-1 \biggr ) \nonumber\\
& +{\sigma_1^\dagger}^m~
\delta \biggl (\sum_{i=1}^N \ad_i a_i-n+N-1 \biggr ).
\label{bosonizeapp}
\end{align}


\bibliography{references.bib}
\bibliographystyle{JHEP} 

\end{document}